\documentclass[useAMS,usenatbib]{mn2e}
\usepackage{epsfig}
\usepackage{natbib}
\usepackage{lscape}
\usepackage{rotating}
\usepackage{amsmath,amssymb}
\usepackage{subfigure}
\DeclareMathVersion{bold}

\def\lsim{\,\lower2truept\hbox{${<\atop\hbox{\raise4truept\hbox{$\sim$}}}$}\,}
\def\gsim{\,\lower2truept\hbox{${> \atop\hbox{\raise4truept\hbox{$\sim$}}}$}\,}

\title[Polarisation of bright AT20G sources]{A Polarisation Survey of Bright Extragalactic \\ AT20G Sources}
\author[Massardi et al.]{\parbox[t]{\textwidth}
{M. Massardi$^{1}$\thanks{Email: massardi@ira.inaf.it}, S. G. Burke-Spolaor$^2$, T. Murphy$^{3,4}$, R. Ricci$^{1}$, M. L\'opez-Caniego$^{5}$, M. Negrello$^{6}$, R. Chhetri$^{7,2}$, G. De Zotti$^{6,8}$, R. D. Ekers$^2$, R. B. Partridge$^{9}$,\\ E. M. Sadler$^{5}$ }
\vspace*{8pt} \\
$^{1}$INAF - Istituto di Radioastronomia, Via Gobetti 101, I-40129, Bologna, Italy\\
$^{2}$Australia Telescope National Facility, CSIRO, P.O. Box 76, Epping, NSW 1710, Australia\\
$^{3}$School of Physics, University of Sydney, NSW 2006, Australia\\
$^{4}$School of Information Technologies, University of Sydney, NSW 2006, Australia\\
$^{5}$Instituto de F\'isica de Cantabria (CSIC-UC), Avda. los Castros s/n, E-39005 Santander, Spain\\
$^{6}$INAF - Osservatorio Astronomico di Padova, vicolo dell'Osservatorio 5, I-35122, Padova, Italy\\
$^{7}$Department of Astrophysics and Optics, School of Physics, University of New South Wales, NSW, 2052, Australia\\
$^{8}$SISSA, Via Bonomea 265, I-34136 Trieste, Italy \\
$^{9}$Haverford College Astronomy Department, 370 Lancaster Avenue, Haverford, PA, 19041 USA\\
}

\begin{document}

\maketitle

\begin{abstract}
We present polarisation data for 180 extragalactic sources extracted from the Australia Telescope 20 GHz (AT20G) survey catalog, and observed with the Australia Telescope Compact Array during a dedicated, high sensitivity run ($\sigma_P\sim 1$\,mJy). For the sake of completeness we extracted the polarisation information for 7 extended sources from the 9-yr WMAP coadded maps at 23 GHz. The full sample of 187 sources constitutes a $\simeq 99\%$ complete sample of extragalactic sources brighter than $S_{20\rm GHz}=500\,$mJy at the selection epoch  with declination $\delta < -30^\circ$. The sample has a $91.4\%$ detection rate in polarisation at $\sim$20~GHz ($94\%$ if considering the sub-sample of point like sources). We have measurements also at 4.8 and 8.6 GHz within $\sim$1 month of the 20~GHz observations for 172 sources to reconstruct the spectral properties of the sample in total intensity and in polarisation: 143 of them have a polarisation detection at all three frequencies. 

We find that there is no statistically significant evidence of a relationship either between the fraction of polarisation and frequency or between the fraction of polarisation and the total intensity flux density. This indicates that Faraday depolarisation is not very important above 4.8 GHz and that the magnetic field is not substantially more ordered in the regions dominating the emission at higher frequencies (up to 20 GHz). We estimate the distribution of the polarisation fraction and the polarised flux density source counts at $\sim$20~GHz.
\end{abstract}


\section{Introduction}\label{sec:intro}

\begin{table*}
 \caption{AT20G and WMAP related catalogues that include data in polarisation. \label{Table:otherpolsurveys}}
 \centering
\resizebox{18cm}{!}{
 \begin{tabular}{lccl}
\hline
\textbf{References}      &\textbf{Frequency(GHz)}&\textbf{\# sources}& \textbf{Notes}   \\
 \hline
Massardi et al. (2008) AT20G-BSS 	& 4.8, 8.6, 20			& 320 & AT20G bright sample\\
L\'opez-Caniego et al. (2009)			& 23,33,41				& 22 & polarisation detection in WMAP map \\
Jackson et al. (2010), Battye et al. (2011) & 8.4, 22, 43 	& 230& WMAP sources\\
Sajina et al. (2011)				& 4.8, 8.4, 22, 43		& 159& equatorial AT20G sources \\
Murphy et al. (2010), Massardi et al. (2011a) AT20G & 4.8, 8.6, 20 & 5890& AT20G 91\% complete sample above 100\,mJy\\
Burke-Spolaor et al. (2009)			& 18			 & 9 & extended sources in the Southern hemisphere \\
current paper 						& 4.8, 8.6, 18			 & 193 & complete sample above 500\,mJy \\
\hline
 \end{tabular}}
 \end{table*}

The study of the properties of radio source populations above 10\,GHz has progressed greatly in recent years, fostered  by Cosmic Microwave Background (CMB) observation campaigns that require an accurate understanding of the contamination of the CMB signal by foreground sources. Extragalactic radio sources are the dominant contaminant on angular scales smaller than 30 arcmin, both in total intensity and in polarisation at frequencies of up to $\simeq 100$--$200\,$GHz (Toffolatti et al. 1998; 1999; De Zotti et al. 1999). An accurate determination of their emission is therefore important to get clean CMB maps and is absolutely critical for the detection of the extremely weak cosmological B-mode polarisation (see Tucci \& Toffolatti 2012 for a recent review).

Extending the characterization of the polarisation properties of radio sources to high frequencies is interesting {\it per s\'e}, as it provides information about the physics of the emission process. In compact, Doppler boosted sources that dominate the high-frequency population at bright flux density levels, the emission at higher and higher frequencies mostly arises from synchrotron, self-absorbed, knot-like structures in the relativistic jet closer and closer to the active nucleus (e.g. Blandford \& K\"onigl 1979). It has been argued that  the ordering of magnetic fields should increase in the inner regions, and as a consequence, the polarisation degree increases (Tucci et al. 2004).

However, the polarisation properties of high-frequency extragalactic contaminants are still poorly constrained by observations. Most current estimates rely on extrapolations from low-frequency samples; the NVSS at 1.4\,GHz (Condon et al. 1998) still constitutes the largest sample of sources surveyed both in total intensity and polarisation. Extrapolations are affected by large uncertainties since a complex combination of effects must be considered. This includes intra-beam effects and bandwidth depolarisation, in addition to intrinsic frequency-dependent changes. The propagation of the radiation through diffuse plasma screens between the source and the observer can cause depolarisation and rotation of the polarisation angle. These effects are difficult to isolate observationally, although we can benefit from the inverse square frequency dependence of the latter effect.

Because the polarised signal in extragalactic objects is typically a few percent of the total intensity signal, deep surveys are necessary to collect statistically significant samples. But high-frequency, deep surveys are time-consuming for diffraction-limited, ground-based telescopes. This has motivated sensitive high frequency polarisation measurements of source samples usually selected from surveys at $\lsim$5\,GHz (Klein et al. 2003; Ricci et al. 2004; Jackson et al. 2007; Agudo et al. 2010; see Tucci et al. 2004 for a summary of earlier polarisation surveys).

The study of polarisation properties of complete samples selected at $\ge 20\,$GHz has become possible thanks to the WMAP all-sky survey with a completeness limit of $\simeq 1\,$Jy at 23 GHz (Arg{\"u}eso, Gonz{\'a}lez-Nuevo,
\& Toffolatti 2003; De Zotti et al. 2005; Wright et al. 2009; Gold et al. 2011) and to the Australia Telescope 20 GHz (AT20G) survey (Murphy et al. 2010; Hancock et al. 2011) that has covered the full southern sky with a 91\% completeness above 100 mJy and 79\% completeness above 50 mJy in regions south of declination $-15^\circ$ (north of $-15^\circ$ the catalogue completeness is lower between 14 and 20 h in right ascension). Table \ref{Table:otherpolsurveys} lists some details on the AT20G and WMAP related samples that include data in polarisation. Larger samples of radio sources selected at higher frequencies are being provided by the  {\sc Planck} mission (Planck 2011a,b, 2013).

Multi-steradian samples of high-frequency selected polarised sources are also important for identifying suitable calibrators for CMB polarimetric experiments and upcoming millimeter-wave telescopes. Pictor A has been identified as a suitable extragalactic polarisation calibrator for the  {\sc Planck} Low Frequency Instrument (LFI) because of its position (in the region of the Ecliptic Pole where the satellite scans $\sim$ once per minute) and its lack of variability from the hot spots that dominate the polarised signal (Burke-Spolaor et al. 2009).

The polarisation of WMAP sources has been investigated by L{\'o}pez-Caniego et al. (2009) using WMAP data; 14 extragalactic sources were significantly detected in polarisation. Follow-up observations of a complete sample of 203 WMAP sources were carried out with the VLA by Jackson et al. (2010); polarised emission was detected for 123, 169 and 167 at 8.4, 22 and 43 GHz respectively.

Sadler et al. (2006) presented polarisation measurements for a sample of 173 AT20G sources brighter than $S_{\rm 20 GHz}=100$\,mJy; 129 ($\simeq 75\%$) were detected at 20 GHz, with a median fractional polarisation of 2.3\%. Massardi et al. (2008) discussed the polarisation properties of the AT20G bright sample ($S_{20\rm GHz}\ge 500\,$mJy), finding 213 $\ge 3\sigma$ polarisation detections at 20 GHz out of a total of 320 sources ($\simeq 67\%$), with a median fractional polarisation of $2.5\%$ at 20 GHz. The spectral indices in total intensity and in polarisation were found to be similar on average, but there were several sources for which the spectral shape of the polarised emission is substantially different from the spectral shape in total intensity. The full AT20G catalog (Murphy et al. 2010; Massardi et al. 2011a) includes the 20 GHz polarised intensity for 768 sources, 467 of which also have simultaneous polarisation detections at 5 and/or 8 GHz, out of a total of 5890 sources. The detection limit is defined as $\max(3\sigma,0.01S_{20\rm GHz},6\,\hbox{mJy})$.

Sajina et al. (2011) obtained polarisation measurements with the VLA at 4.86, 8.46, 22.46, and 43.34 GHz of 159 out of the $\simeq 200$ AT20G radio galaxies with $S_{20\rm GHz}\ge 40\,$mJy in an equatorial field of the Atacama Cosmology Telescope survey; polarised flux was detected at $> 95\%$ confidence level for 141, 146, 89, and 59 sources, from low to high frequencies. The measured polarisation fractions are typically $<5\%$, although in some cases they are measured to be up to $\simeq 20\%$. For sources with detected polarised flux in all four bands, about 40\% of the sample, the polarisation fractions typically increase with frequency. This trend is stronger for steeper spectrum sources as well as for the lower flux density sources.

The conclusions of all the polarisation studies of complete samples selected at high frequencies are limited by the moderate detection rates. To overcome this limitation we have performed dedicated high-sensitivity polarisation observations of a complete AT20G bright source sub-sample achieving a $\simeq 93\%$ detection rate. In addition to allowing a more thorough investigation of the polarisation properties of the high-frequency radio source populations, this sample constitutes a legacy data set for polarisation studies in the Southern hemisphere. 
In particular, this sample could help in the definition of calibrator source lists for facilities working in the millimetric bands, like the Atacama Large Millimeter Array (ALMA) that, in its final configuration, will observe from 30 to 950\,GHz in both total intensity and polarisation.

This paper is structured as follows. In \S\,\ref{sec:selection} we describe the sample selection and the observing strategy, and in \S\,\ref{sec:reduction} the data reduction techniques. For the sake of completeness, we included the polarised flux density measurements in this analysis as well as limits for the extended sources as extracted from low resolution 9-year WMAP maps described in \S\, \ref{sec:WMAP}. The final sample thus includes 187 sources. In \S\,\ref{sec:analysis} we present the analysis of their spectral and polarisation properties. In \S\,\ref{sec:counts} we derive the source counts as a function of the polarised flux density. In \S\,\ref{sec:cal} we present a selection of candidate polarisation calibrators for millimetric wavelength studies.
The main conclusions are summarised in \S\,\ref{sec:conclusions}.

\section{Source selection and observations}\label{sec:selection}
The selection of the sample was based on the list of confirmed AT20G sources available at the epoch of our observations (October 2006). We selected all objects with flux density $S_{20{\rm GHz}}>500\,$mJy and declination $\delta<-30^\circ$, excluding the Galactic plane region ($|b|\le 1.5^\circ$) and the Large Magellanic Cloud Region (inside a circle of $5.5^\circ$ radius centered at $\alpha=05:23:34.7$ and $\delta=-69:45:22$). This resulted in a complete sample of 189 sources.

Nine of them were found, with the aid of low frequency radio imaging surveys (PMN, Griffith \& Wright 1994, 1995, and SUMSS, Mauch et al. 2003), to be very extended. These were observed with the mosaic mode at 20\,GHz by Burke-Spolaor et al. (2009). Flux densities integrated over the whole source are available for 5 of them. For the remaining 4 objects the measured integrated flux densities, if available, refer to subregions. For these 4 objects we decided to extract the integrated flux densities from the low resolution WMAP maps as described in Sect. \ref{sec:WMAP}.

It should be noted that the flux densities reported in the final published AT20G catalogue may be slightly different from those given in the preliminary 2006 version used for our source selection. The reason is that in the case of sources observed more than once (which is most likely the case for bright sources such as those considered here) the highest quality observation was listed in the final AT20G catalogue, as discussed by Murphy et al. (2010). Because of variability this has the effect of moving some sources above or below the adopted flux density threshold ($S_{20\rm GHz}=500\,$mJy). Since there are more sources below than above the threshold, there are more sources moving up than moving down, and we end up with slightly more sources above threshold in the final catalogue than in the version we have used. Hence there are 214 sources listed in the final AT20G catalogue with declination $\delta<-30^\circ$ and flux density above our chosen threshold. 

On the other hand, of the 180 sources with good quality flux density in our ATCA observations (as discussed below) only 165 still have $S_{20\rm GHz}\ge 500\,$mJy in the final catalogue and only 145 were found to be above this threshold in our October 2006 observations. However, despite variability, the final sample, which is reasonably complete at the selection epoch, is representative of the bright 20 GHz population as a whole and can be used to assess statistical properties of this population.

\begin{table*}
 \caption{Catalogue of the first 10 sources of the extragalactic sample observed in the AT20G run dedicated to high sensitivity polarisation. The full source list, available as Supplementary Material online, includes 187 objects.
Columns are: (1) AT20G name; (2) 18\,GHz scalar average flux density (error given by eq.~(\ref{eq:sigi})); (3--5) polarised flux density, fractional polarisation, and polarisation angle at 18\,GHz; (6--9) total intensity, polarised flux density, fractional polarisation and polarisation angle at 8.6\,GHz; (10--13) total intensity and polarised flux densities, polarisation degree and polarisation angle at 4.8\,GHz; (14) ratio between the visibilities amplitude averaged over the long baselines ($\sim4.5$ km) and the short ones ($\sim0.2$ km) in the AT20G catalogue (Chhetri et al. 2013); (15) flags: `s' for sources with low declination observed during the AT20G 20\,GHz run (i.e. the highest frequency is 20 GHz instead of 18 GHz),`f' for sources with data at 20 GHz observed in March 2006 because of the poor quality of their data in the 2006 October runs, `b' for very extended sources included in the Burke-Spolaor et al. (2009) sample, `w' for extended sources which have had flux densities (or their upper limits) extracted from the WMAP 23 GHz 9-yr coadded maps (i.e. the highest frequency is 23 GHz instead of 18 GHz), and `c' for sources candidate to be suitable polarisation calibrators at mm and submm wavelengths.}\label{Table:fullTable}

\resizebox{18cm}{!} {
\begin{tabular}{lcccccccccccccl}
\hline
Name `AT20G'& $S_{\rm 18}$ & $P_{\rm 18}$ & $\Pi_{\rm 18}$ & $\phi_{\rm 18}$ &$S_{\rm 8.6}$ & $P_{\rm 8.6}$ & $\Pi_{\rm 8.6}$ & $\phi_{\rm 8.6}$ &$S_{\rm 4.8}$ & $P_{\rm 4.8}$ & $\Pi_{\rm 4.8}$ & $\phi_{\rm 4.8}$ & 6~km Vis & Flags\\
 & [mJy]&[mJy]& [\%]&[deg] &[mJy]&[mJy]& [\%]&[deg] &[mJy]&[mJy] & [\%]&[deg]&& \\
\hline
 J000435-473619&  798.4\tiny{$\pm$  16.0}&  7.09\tiny{$\pm$ 0.4}&  0.89\tiny{$\pm$ 0.06}& -78.70&  929.0\tiny{$\pm$   0.7}&  17.60\tiny{$\pm$ 0.8}&  1.90\tiny{$\pm$ 0.01}& -51.50&  865.0\tiny{$\pm$   0.6}&  15.70\tiny{$\pm$ 0.7}&  1.82\tiny{$\pm$ 0.01}& -37.04 &0.98 & ...\\
 J001035-302748&  595.8\tiny{$\pm$  11.9}&  31.16\tiny{$\pm$ 1.3}&  5.23\tiny{$\pm$ 0.24}& -46.66&  676.0\tiny{$\pm$   0.7}&  25.40\tiny{$\pm$ 0.9}&  3.75\tiny{$\pm$ 0.01}& -38.97&  582.0\tiny{$\pm$   0.6}&  6.40\tiny{$\pm$ 0.6}&  1.10\tiny{$\pm$ 0.01}& -52.22 &0.96 & c..\\
 J001259-395426& 1030.0\tiny{$\pm$  20.6}&  33.67\tiny{$\pm$ 0.9}&  3.27\tiny{$\pm$ 0.11}& -86.42& 1622.0\tiny{$\pm$   0.8}&  24.70\tiny{$\pm$ 0.8}&  1.52\tiny{$\pm$ 0.01}& -85.56& 1651.0\tiny{$\pm$   0.6}&  20.60\tiny{$\pm$ 0.6}&  1.25\tiny{$\pm$ 0.01}& -88.13&0.96 & ...\\
J002616-351249& 1136.0\tiny{$\pm$  22.8}&  1.65\tiny{$\pm$ 0.4}&  0.15\tiny{$\pm$ 0.04}& -54.21&  382.0\tiny{$\pm$   0.8}& $<$2.40&...&...&  133.0\tiny{$\pm$   0.7}& $<$2.09&...&...                                                                                &0.98 & ...\\
 J004959-573827& 1859.0\tiny{$\pm$  37.2}&  71.41\tiny{$\pm$ 1.6}&  3.84\tiny{$\pm$ 0.11}&  35.39& 2460.0\tiny{$\pm$   0.8}&  67.10\tiny{$\pm$ 0.9}&  2.73\tiny{$\pm$ 0.01}&  9.14& 2237.0\tiny{$\pm$   0.7}&  67.00\tiny{$\pm$ 0.7}&  2.99\tiny{$\pm$ 0.01}&  5.78  &0.97 & c..\\
J005109-422632&  683.7\tiny{$\pm$  13.7}&  11.11\tiny{$\pm$ 0.5}&  1.62\tiny{$\pm$ 0.08}&  -2.21&  950.0\tiny{$\pm$   0.7}&  10.00\tiny{$\pm$ 0.7}&  1.31\tiny{$\pm$ 0.07}&  8.64& 1060.0\tiny{$\pm$   0.6}&  20.00\tiny{$\pm$ 0.6}&  1.46\tiny{$\pm$ 0.06}&  10.43  &...  & ...\\
 J005846-565911&  785.1\tiny{$\pm$  15.7}&  12.60\tiny{$\pm$ 0.6}&  1.60\tiny{$\pm$ 0.09}& -23.19&  709.0\tiny{$\pm$   0.6}&  16.70\tiny{$\pm$ 0.6}&  2.36\tiny{$\pm$ 0.01}& -72.10&  573.0\tiny{$\pm$   0.6}& $<$1.80&...&...                                       &0.99 & ...\\
J010645-403419& 3785.0\tiny{$\pm$  75.7}&  78.84\tiny{$\pm$ 2.9}&  2.08\tiny{$\pm$ 0.09}&  -8.26& 3680.0\tiny{$\pm$   1.5}&  60.00\tiny{$\pm$ 1.5}&  1.70\tiny{$\pm$ 0.04}&  46.43& 2410.0\tiny{$\pm$   0.7}&  30.00\tiny{$\pm$ 0.7}&  1.32\tiny{$\pm$ 0.03}& -62.00 &...  & ...\\
 J010915-604948&  375.0\tiny{$\pm$   7.5}&  15.73\tiny{$\pm$ 0.6}&  4.20\tiny{$\pm$ 0.18}& -14.55&  516.0\tiny{$\pm$   0.7}&  7.40\tiny{$\pm$ 0.7}&  1.44\tiny{$\pm$ 0.01}& -49.53&  533.0\tiny{$\pm$   0.6}&  9.10\tiny{$\pm$ 0.6}&  1.70\tiny{$\pm$ 0.01}&  88.12  &0.97 & ...\\
 J012457-511316&  369.5\tiny{$\pm$   7.4}&  7.00\tiny{$\pm$ 0.5}&  1.90\tiny{$\pm$ 0.13}& -11.21&  298.0\tiny{$\pm$   0.7}&  9.60\tiny{$\pm$ 0.7}&  3.20\tiny{$\pm$ 0.01}& -15.13&  220.0\tiny{$\pm$   0.7}&  2.40\tiny{$\pm$ 0.7}&  1.11\tiny{$\pm$ 0.01}& -31.11   &0.99 & ...\\
 \hline
  \end{tabular}
  }
 \end{table*}
\subsection{Observations in the 20 GHz band}
Observations were taken on October 1, 2006 using the most compact hybrid configuration of ATCA, ``H75'', excluding the data from the farthest antenna. The longest baseline of this configuration is 75\,m, and its T-shape ensures adequate Fourier coverage for snapshots taken on a relatively small range of hour angles and at high elevation.

Although the linear feeds of the ATCA somewhat complicate the polarisation calibration procedure, the array has several inherent advantages for polarisation experiments. The on-axis receivers of the telescope introduce relatively low amounts of instrumental polarisation, while all antennae are fitted with a noise diode that injects a signal to continually track the phase difference (xy-phase) between the two orthogonal feeds. In addition, since the feeds are linearly polarised, there is very little contamination of the circular polarisation signal by the total intensity signal. For further details on the ATCA instrumental polarisation we refer readers to Sault, Reyner \& Kesteven (2002).

A digital correlator (later replaced by the CABB broad-band digital correlator) allowed simultaneous observations in two bands, each with a bandwidth of 128 MHz divided into 32 channels. The observation frequencies within the 20\,GHz band, covering the 16-24\,GHz range, were chosen so as to maximize sensitivity, make use of optimal system temperatures, and avoid correlator harmonics. The frequency bands were centered at 16.704 and 19.392\,GHz. After calibration, data were averaged over the 256 MHz band so that our mean effective frequency was 18.048\,GHz (hereafter tagged as ``18''GHz observations). The FOV was $\sim 2.6$ arcmin.

The closeness of dishes in the H75 configuration can cause significant antenna shadowing for sources south of a declination of approximately $-76^{\circ}$. Because the effects of shadowing and cross-talk on polarisation measurements are unknown for the instrument, the sample for H75 observations was restricted to $-76^{\circ} < \delta < -30^{\circ}$. Eleven $\delta<-76^{\circ}$ sources were observed for this project during a run that took place on 17th October 2006 with a more extended hybrid antenna configuration (H214) and observation bands centered at 18.752 and 21.056\,GHz. Calibration and observation setup for these objects matched that of AT20G follow-up observations (see Murphy et al. 2010 for details). After the data reduction, the data for 3 of these objects (AT20GJ115253-834410, AT20GJ122454-831310, AT20GJ155059-825807) at 20 GHz were found to be of poor quality in this run. However, they had good quality observations and a polarisation detection in March 2006 during a previous AT20G follow-up run. Hence, we decided to include these measurements in the present analysis, flagging them in the catalogue.

For each run a bright point-like source was observed to calculate bandpass solutions and PKS\,1934--638 was used as the primary flux density calibrator. The sample was broken into groups of 4--6 sources, located in the same sky region. Before each group was observed, an antenna pointing correction was performed using a nearby bright source to maintain directional accuracy despite the vast range of sky positions that the telescope had to observe in a short period of time. Each target was observed in two 70-second snapshots separated by four hours. 

Five targets were identified as extended within the primary beam, according to the extendedness criteria described in Murphy et al. (2010). As the flux density estimation methods that we use (see Sect.\ref{sec:reduction}) are well suited for compact objects, the flux densities of extended sources are likely underestimated by an unknown amount and the polarised flux density is potentially wrong. For this reason they were included in the list of extended sources for which we extracted the polarisation information from the WMAP 9-yr coadded maps (see Sect. \ref{sec:WMAP}). In the end, after dropping 4 of the 9 very extended sources observed by Burke-Spolaor et al. (2009) and the 5 sources found to be extended, we are left with 180 sources for which we have good quality ATCA data.

Five antennae were available in our 16.704 GHz band, while there was an error in the 19.392\,GHz band that left only four usable antennae at that frequency. To save observing time, no secondary calibrators were interleaved with the target observations, with the intention of self-calibrating each bright source during the data reduction. As described in Sect. \ref{sec:reduction}, this turned out to be an unwise strategy in terms of polarisation calibration. Nevertheless, 169 of the 180 extragalactic sources in our ATCA observed sample (94\%) were detected in polarisation at 18 GHz. However, only 145 of them also have $S_{18\rm GHz}>500$ mJy  in the October 2006 observations.

Table~\ref{Table:fullTable} provides an excerpt of the catalogue containing the data for the sample of 180 objects for which we have good quality ATCA data.

\subsection{Lower-frequency observations}
Lower-frequency observations were performed during a regular AT20G follow-up run (described in Murphy et al.\, 2010), carried out in November 2006. In that epoch, we used a 5-antenna east-west array configuration, with a 1.5\,km longest baseline. Two 30-second observations per source provided simultaneous 4800 and 8640\,MHz snapshots. These observations had 128 MHz of bandwidth per frequency. The FOV were of 9.9 and 5.5 arcmin, respectively. 

Massardi et al. (2011b) demonstrated that the median variability of total intensity flux densities for sources in flux density ranges similar to those of our sample over 3 month timescale is 3.5 and 6.3 per cent respectively at $\sim 5$ and $\sim 9$ GHz, sligthly increasing with time lag. Kurinski et al. (2013) found a median variability index for total intensity flux densities of 1.0\% at 5 GHz on less than 2 months time lags for a similar sample in the Northern hemisphere. Hence we could assume that our high and low frequency observation comparisons, on shorter time scales, are not substantially affected by variability for most sources.

173 and 172 sources have good quality flux densities respectively at 4.8 and 8.6 GHz. 172 sources have good quality flux densities at all the frequencies (quality controls are described in Murphy et al.\, 2010). Of these, 137 have $S_{18\rm GHz}>500$ mJy in the October 2006 run. 143 sources have a polarisation detection at all the frequencies (79\% of the main sample), 119 of which have $S_{18\rm GHz}>500$ mJy in the October 2006 run.

\section{Data Reduction} \label{sec:reduction}
The 16.7 and 19.4\,GHz data were reduced using the {\sc Miriad} software package (Sault, Teuben, \& Wright 1995). The two frequencies were calibrated independently and then combined for 18\,GHz imaging and flux density assessment. Opacity correction and a correction for the time-dependent instrumental xy-phase difference was applied upon loading all data into {\sc Miriad}. After this correction, a small residual gain offset still remained to be corrected in the following calibration stages. Bandpass solutions and primary flux density calibration were calculated and applied using PKS\,1921--293 and PKS\,1934--638, respectively.

For polarimetric calibration with calibrators of unknown polarisation and sparse data (such as in our short snapshot observations), the standard {\sc Miriad} procedure suggests calculating the largely stable instrumental leakage terms by using an unpolarised primary calibrator. The remaining polarisation and gain terms are then calculated for each secondary phase calibrator. 

Roughly 75\% of the sources in our sample are registered in the ATCA calibrator database; all are sufficiently bright to determine adequate calibration solutions. However, though it is suggested that accurate Q and U values could be calculated from a relatively small amount of data, it was apparent that this was not the case for our $\sim$3 minute observations. Many solutions failed, while others produced impossible values for Q and U levels. This is likely due to the large number of free parameters and insufficient data length, even using the smallest possible solution interval. To overcome this hindrance, all polarisation and gain solutions for the main sample were determined using the secondary calibrators that were interlaced in our observations, though these were originally intended for use with the objects presented by Burke-Spolaor et al. (2009). Merged solutions from the eight available calibrators afforded an observation every 1-2 hours in various regions of the sky. Because each was typically observed at least for one minute at any of eight parallactic angle intervals spanning approximately 6 hours, they had sufficient parallactic angle coverage to disentangle instrumental polarisation from the intrinsic calibrator Stokes Q and U levels. The absolute flux density scale calculated from PKS\,1934--638 was then used to bootstrap the secondary calibrators; gain and instrumental polarisation terms were then applied to all target sources.

Incidentally, tests run on the calibrated sources to check levels of residual phase instability, which is usually due to imperfect phase calibration (as given in percentage by dividing the source vector amplitude by its scalar amplitude) showed that using calibrators observed frequently in time can give a better phase calibration than less frequently observed calibrators even if they are  closer in space to the target source (as shown by 0-20\% residual decorrelation in our sample versus the 0-50\% found in AT20G data over similar time scales and weather conditions). 

Hence, in polarimetry experiments covering large areas of the sky, it appears more pertinent to have many observations of one calibrator throughout an observation and therefore have sufficient data to determine accurate polarisation solutions, despite possible non-proximity to target sources. However, calibrators in this experiment were never further than $30^{\circ}$ from any target source and were restricted to high elevation observations.

The 4.8 and 8.6\,GHz data for the main sample and the 20\,GHz data for the 11 $\delta < -76^{\circ}$ sources had observational modes exactly coincident with the AT20G survey follow-up, and thus were flagged and reduced using the automatic pipeline developed for the AT20G (Murphy et al. 2010).

Stokes-I intensities were determined from the visibilities to avoid the inclusion of phase instabilities inherent in image-based measurements. This technique takes the scalar average of the visibility amplitudes and is robust for bright ($>$200\,mJy), point-like sources only.

To acquire Stokes $Q$, $U$, and $V$ flux densities, images were created and deconvolved using the {\sc Miriad} task CLEAN. To correct the Stokes $Q$, $U$, and $V$ images for decorrelation, we took advantage of the fact that Stokes parameters, simultaneously measured, are affected by decorrelation originating in atmospheric phase instabilities (as might be left after imperfect calibration). We can thus use the fractional level of residual decorrelation ($\chi$) in Stokes $I$, calculated and applied to $Q$, $U$, and $V$ flux densities as:
\begin{equation}
\chi = \frac{I_{\rm sca}}{I_{\rm map}}
\label{eq:decorr}
\end{equation}
\begin{equation}
Z = \chi \cdot Z_{\rm map},
\label{eq:decorrcorr}
\end{equation}
where \textit{Z} represents Stokes $Q$, $U$ or $V$, $I_{\rm sca}$ is the scalar-averaged Stokes I flux density, and $I_{\rm map}$  and \textit{Z}$_{\rm map}$ represent the values at the position of the peak Stokes I emission in the relevant image. The image peak for all sources was sufficient to determine the decorrelated flux density measurements; the pixel size was typically 10\,arcsec, and no sources in this subsample were extended significantly beyond this.

The polarised intensity and the position angles were then calculated using standard first-order debiasing, where $P = \sqrt{Q^2 + U^2 - \sigma_{\rm V}^{2}}$ (Wardle \& Kronberg 1974; Simmons \& Stewart 1985). The last term, $\sigma_{\rm V}$, is the RMS noise in the Stokes $V$ image. Most extragalactic sources do not have significant levels of circular polarisation; therefore the Stokes $V$ signal is usually undetected or very low for all such sources. This point makes the noise level in Stokes $V$ a reasonable estimate of the background noise level in the $Q$ and $U$ images, and thus gives a good estimate for the debiasing correction.

The polarisation angles are given by $\phi = (1/2)\arctan{(U/Q)}$, and the linear polarisation degree, $\Pi$, is given in the percentage: $\Pi = 100\cdot P/I$.

\begin{table*}
\centering
 \caption{Polarised and total flux densities measurements (in mJy) for the 3 extended sources with a polarisation detection in the WMAP 9-year co-added full-sky maps.}
\label{Table:wmap}
\resizebox{18cm}{!} {
\begin{tabular}{lcccccccc}
\hline
Name & $P_{\rm 23 GHz}$& $P_{\rm 33 GHz}$& $P_{\rm 41 GHz}$ &$P_{\rm 64 GHz}$ & $S_{\rm 23 GHz}$& $S_{\rm 33 GHz}$& $S_{\rm 41 GHz}$ &$S_{\rm 64 GHz}$\\
\hline
Fornax A (RA:03:22:41.7; DEC:-37:12:30)     & 1074 \tiny{31}& 867 \tiny{44} &  589 \tiny{64} &  $<$354 &  9321  \tiny{134}& 	 5350 \tiny{184}& 	 3275 \tiny{173}& 	 905 \tiny{255} \\
PicA - AT20GJ051949-454643&  457 \tiny{35}& 372 \tiny{50} &280 \tiny{82}   & 484 \tiny{137}& 6464 \tiny{207}& 	 5661 \tiny{235}& 	 4656 \tiny{210}& 	 3139 \tiny{270} \\
CenA - AT20GJ132527-430104& 3322 \tiny{70}&2699 \tiny{81}&  2323 \tiny{120}&  2075 \tiny{173}& 51006  \tiny{260}& 	41909 \tiny{248}& 	35731 \tiny{245}& 	26767 \tiny{335} \\
\hline
 \end{tabular}
 }

\end{table*}
\subsection{Error budget}
The RMS scatter $\sigma_{\rm V}$ provided a measurement of the noise on the scalar average flux density of the order of 1-2\%. Telescope pointing inaccuracies, considering the $\sim$15 arcsec pointing errors and the 18\,GHz primary beam response function are expected to cause a possible attenuation of up to 2\% in all Stokes parameters. 

Errors in the primary flux density scaling from PKS\,1934--638 are estimated by comparing the online ATCA calibrator catalogue data at 18.496\,GHz to the high-frequency polynomial model used by {\sc Miriad} to calculate the scaling factor. The model predicts $I_{\rm model} = 1.0259\,$Jy at our observing frequency while the average of values measured in 2006 is $I_{\rm data} = 1.0278\,$Jy; the calibration error in source flux densities is thus approximately 0.2\%. Given the much larger statistical errors in $I_{\rm sca}$, we ignore this calibration error. The net error in total intensity is thus given by:
\begin{equation}\label{eq:sigi}
\sigma_I^2 = (0.02\,I_{\rm sca})^2 + \sigma_{\rm V}^2.
\end{equation}

Errors in the Stokes parameters $Q$, $U$, and $V$ can arise from leakage of the much brighter Stokes $I$ signal due to imperfections in the alignment of orthogonal receiver components. The correction of this effect using PKS\,1934--634 and an iterative leakage calculation using secondary calibrators results in a negligible error compared to the system noise. The noise term is calculated by propagation of errors through eqs.~(\ref{eq:decorr}) and (\ref{eq:decorrcorr}). The main contributions to the global error come from the antenna pointing inaccuracy and from the noise estimated by the rms levels in off-source regions in the restored image:
\begin{equation}\label{eq:stokeserr}
\centering
\sigma_{Z}^2 = (\frac{I_{\rm sca}\sigma_{Z\rm map}}{I_{\rm map}})^2 + (\frac{ZI_{\rm sca}\sigma_{I\rm map}}{I_{\rm map}})^2 + (0.02Z)^2,
\end{equation}
where $Z$ is either $Q$ or $U$. The error on the polarised intensity can then be derived as:
\begin{equation}\label{eq:polerr}
\sigma_P^2 = \frac{Q^2\sigma_Q^2 + U^2\sigma_U^2}{Q^2 + U^2},
\end{equation}
where $Q$ and $U$ are calculated from eq.~(\ref{eq:decorrcorr}). Note that if $\sigma_Q \simeq \sigma_U$ (as expected for low polarisation sources in noisy maps), then $\sigma_P = \sigma_Q = \sigma_U$. We defined as a non-detection of a source $P < 3\,\sigma_P$, and used $3\,\sigma_P$ as the upper limit on the polarised flux density for such sources. The errors in the fractional polarisation were obtained with the usual error propagation from $\sigma_P$ and $\sigma_I$.

\section{Polarisation of extended sources}\label{sec:WMAP}

As mentioned above, our ATCA data on extended sources are largely incomplete. Of the 9 very extended sources with $S_{20\rm GHz}>0.5\,$Jy selected by Burke-Spolaor et al. (2009) for wide field imaging and polarimetry, one (Fornax\,A) was not observed due to its highly diffuse emission, most of which would remain undetected even in ATCA's most compact configuration, and for 3 more (AT20GJ013357-362935, AT20GJ132527-430104, AT20GJ133639-335756) only subregions containing compact structure were observed. Moreover 5 objects in our initial sample were found to be extended with respect to our synthesized beam and therefore to have unreliable polarised flux density estimates in the AT20G catalogues.

For the sake of completeness we have attempted to estimate the polarised flux densities of these extended objects using the WMAP 23 GHz 9 year coadded map, where the 0.88 degree beam is collecting most of the extended emission seen by ATCA. 

The extraction of the total intensity flux density was performed using the IFCAMEX software package\footnote{http://max.ifca.unican.es/IFCAMEX}, which has been used in the past to extract flux densities from WMAP (Massardi et al. 2009) and \textit{{\sc Planck}} data (Planck Collaboration 2013 results XXVIII). The extraction of polarised flux densities from the WMAP 9 year data has been performed using the IFCAPol software package used to characterize polarised sources in WMAP 5 year maps (L\'opez-Caniego et al. 2009). This software implements the Filtered Fusion approach (Arg\"ueso et al. 2011), where a maximum likelihood estimator is obtained for the $Q$ and $U$ maps of each source. As a result, de-noised $Q_{f}$ and $U_{f}$ maps are produced and the polarised flux density at the position of the source is obtained from the map of $P=\sqrt{Q_{f}^2 + U_{f}^2}$.

Note that the WMAP polarisation maps are very noisy and it is important to assess whether or not our estimate of the polarised flux density at the position of a source detected in the total intensity comes from the source or from a maximum of the CMB at that position. This is done by assessing the significance of each detection/estimation in the P map. For each source, we calculate the $99.90\%$ significance level, as explained in L\'opez-Caniego et al. (2009), and check that our estimate of the flux density at the position of the source is at least above this level. This allows us to discriminate between truly significant detections in the maps of $P$ from random peaks of the background.

Suitable detections were obtained at 23 GHz for 2 of the 9 sources in the above mentioned sub-sample (Fornax\,A and Centaurus A). Extractions allowed us to define an upper limit of the integrated polarised flux density for the other 7 cases, but for 2 of these objects the extraction algorithm could not determine the total flux density. 

The results of this extraction are listed in the main catalogue (see Table~\ref{Table:fullTable}), flagged with `w'. The 23 GHz flux densities for the extended sources will be included in the following analysis without any correction for the spectral behaviour between the WMAP observing frequency and the 18 GHz band. Hence, the full sample that will be used in the next section includes 187 sources; it is $99\%$ complete with $S_{20 GHz}>500$mJy at the 2006 survey selection epoch. The polarised emission detection rate is $91.4\%$.

Fornax A is one of the closest and most extended sources in the
Southern Hemisphere, with two lobes extended over a region more than 50 arcmin wide. Its weak core has a flux density at 20 GHz much below the AT20G survey 10 mJy detection limit, and for this reason it is not included in the AT20G catalogue. WMAP was able to integrate over the source area only at 23 GHz, which we included in our analysis. 

In addition, a detection of polarised flux density at 23 GHz in the WMAP maps has also been obtained for Pictor A. It is among the targets observed in mosaic mode by Burke-Spolaor et al. (2009), where it was identified as the best polarisation calibrator among the extragalactic sources for arcmin resolution experiments (and in particular for the  {\sc Planck} experiment), despite its steep spectrum in the region $\sim1-20$ GHz. The detected value of $457\pm 35$ mJy at 23 GHz (listed in Table \ref{Table:wmap}) is comparable with the $500\pm 60$ mJy measured for polarised flux density at 18 GHz by Burke-Spolaor et al. (2009) over the whole source (listed in Table \ref{Table:fullTable} and used in the following analysis). The WMAP detections seem to indicate a steep spectrum in polarised emission in the WMAP frequency range.

Table \ref{Table:wmap} lists the flux densities in total intensity and polarisation for ForA, PicA, and CenA in all the WMAP bands. Notice that the detections at frequencies above 23 GHz might refer only to fractions of the sources if they are more extended than the WMAP beams.

\section{Data Analysis}\label{sec:analysis}

\begin{figure}
 \includegraphics[trim=0cm 0cm 1cm 2cm, height=10cm, clip=true, angle=90]{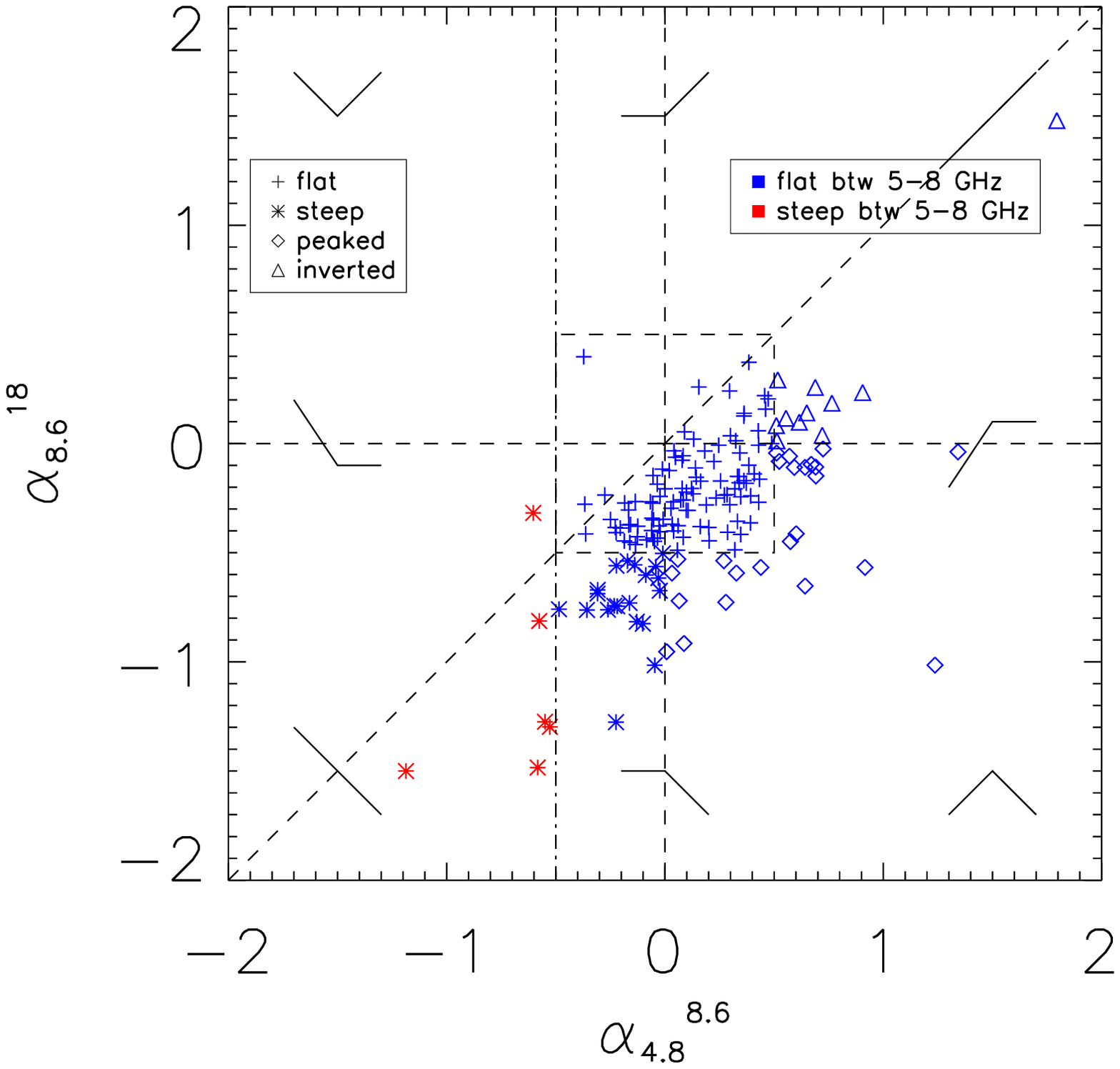}
 \includegraphics[trim=0cm 0cm 1cm 2cm, height=10cm, clip=true, angle=90]{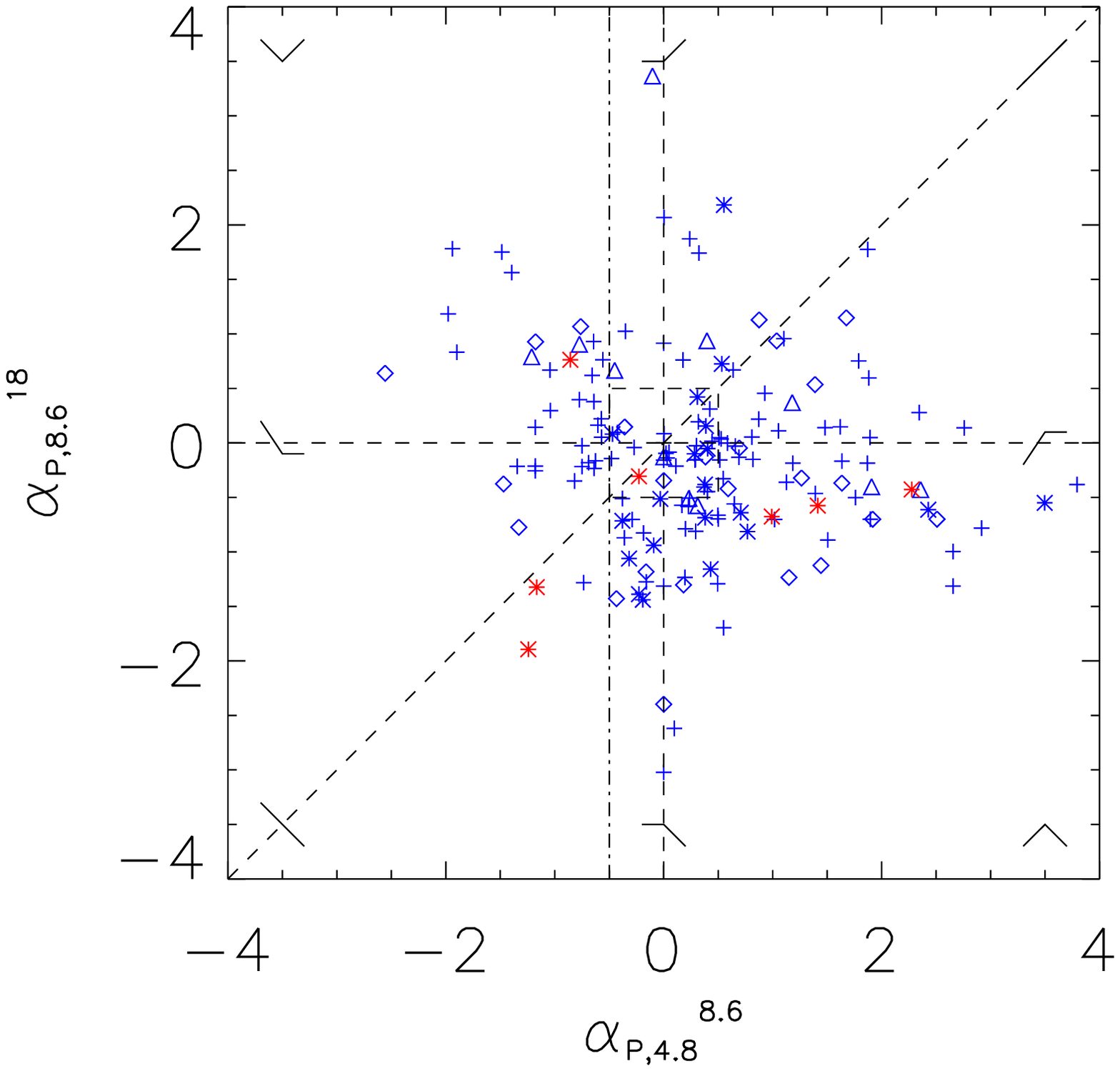}
 \caption{Radio colour-colour diagram for (from top to bottom) flux density and polarised flux density. Error bars and upper limits have been omitted for clarity of display.}
 \label{fig:cc}
\end{figure}

\begin{table}
 \caption{Matrix of the number of sources classified according to both the total intensity and polarisation spectral behaviour. The columns are the spectral shape in polarisation, the rows the spectral shape in total intensity.
 The spectral types are defined in the text.} \label{Table:matrix}
 \resizebox{8cm}{!}{
  \begin{tabular}{|l|c|c|c|c|c|}
 \hline
 {S $\rightarrow$}&(I)&(P)&(F)&(S)&(U)\\
 {Pol. $\downarrow$ } &&&&&\\
 \hline
 Inverted (I) &2&4&21&2&0\\
 Peaked (P) &5 &10 & 33 & 10 & 0 \\
 Flat (F) &1& 3 & 17 & 7 & 1 \\
 Steep (S) & 0&4 & 15 & 9 & 0 \\
 Upturning (U) &4 &3 & 17 & 1 & 0 \\
 \hline
\end{tabular}}
\end{table}

\begin{table}
\centering
 \caption{Median values of spectral indices in different frequency ranges in total intensity and polarisation. Spectral classes are selected according to their behaviour between 4.8 and 8.6 GHz.}
\label{Table:spind}
 \resizebox{8cm}{!}{
\begin{tabular}{ccccc}
\hline
 &    & 4.8-8.6 GHz & 8.6-18 GHz& 4.8-18 GHz \\
\hline
$\alpha_{\nu1}^{\nu2}$ & All  & 0.09 & -0.28 & -0.11\\
            & Flat  & 0.12 & -0.27 & -0.08\\
            & Steep & -0.60 & -1.27 & -0.95\\
\hline
$\alpha_{P,\nu1}^{\nu2}$& All  & 0.20& -0.16 & -0.006\\
            & Flat  & 0.23& -0.15& -0.006\\
            & Steep &-0.23& -0.43& 0.04\\
\hline
\end{tabular}}
 \end{table}
\subsection{Spectral properties of the sample}\label{sec:spectra}
\begin{figure}
 \includegraphics[trim=0cm 0cm 1cm 2cm, height=9cm, clip=true, angle=90]{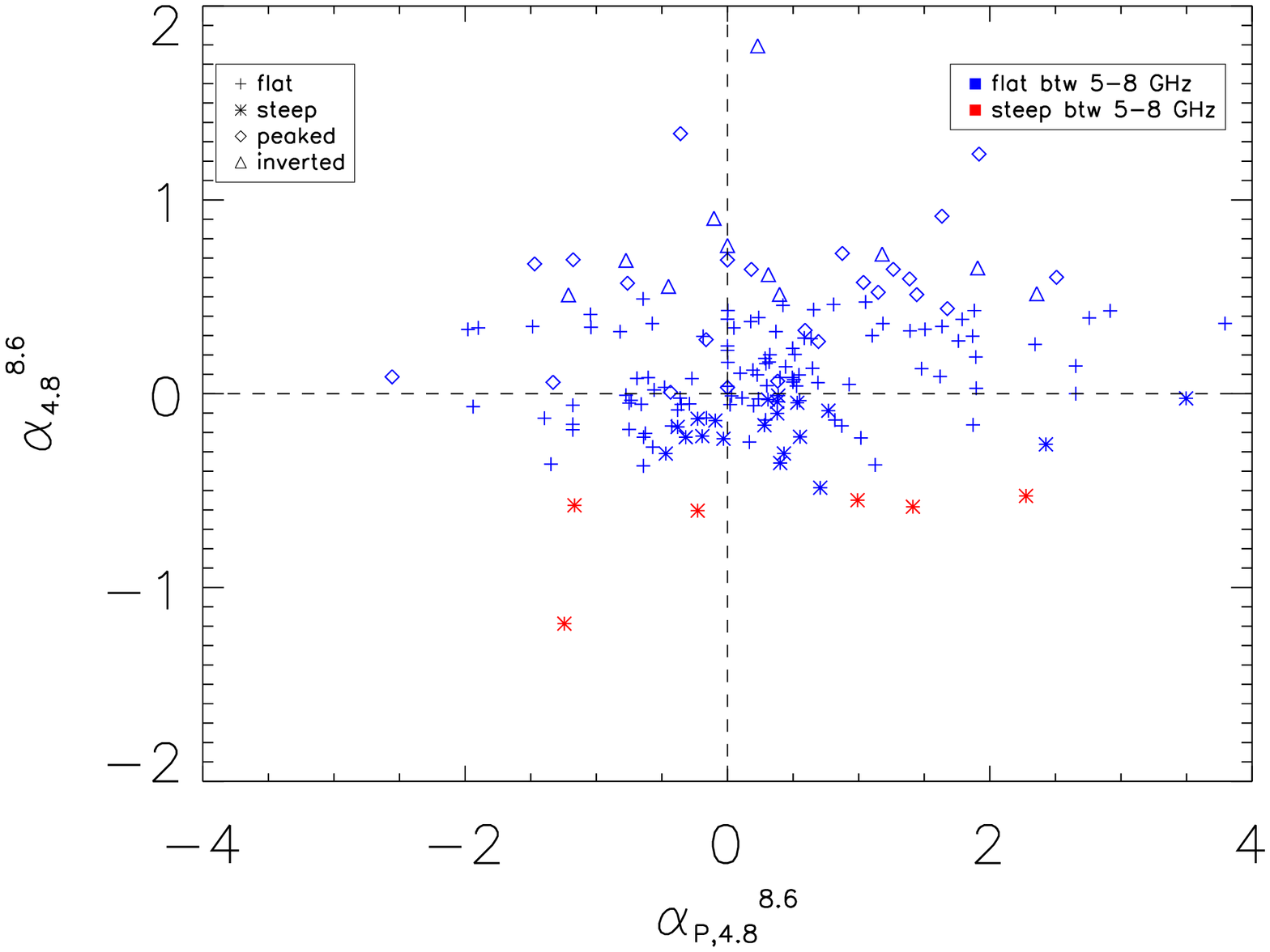}
 \includegraphics[trim=0cm 0cm 1cm 2cm, height=9cm, clip=true, angle=90]{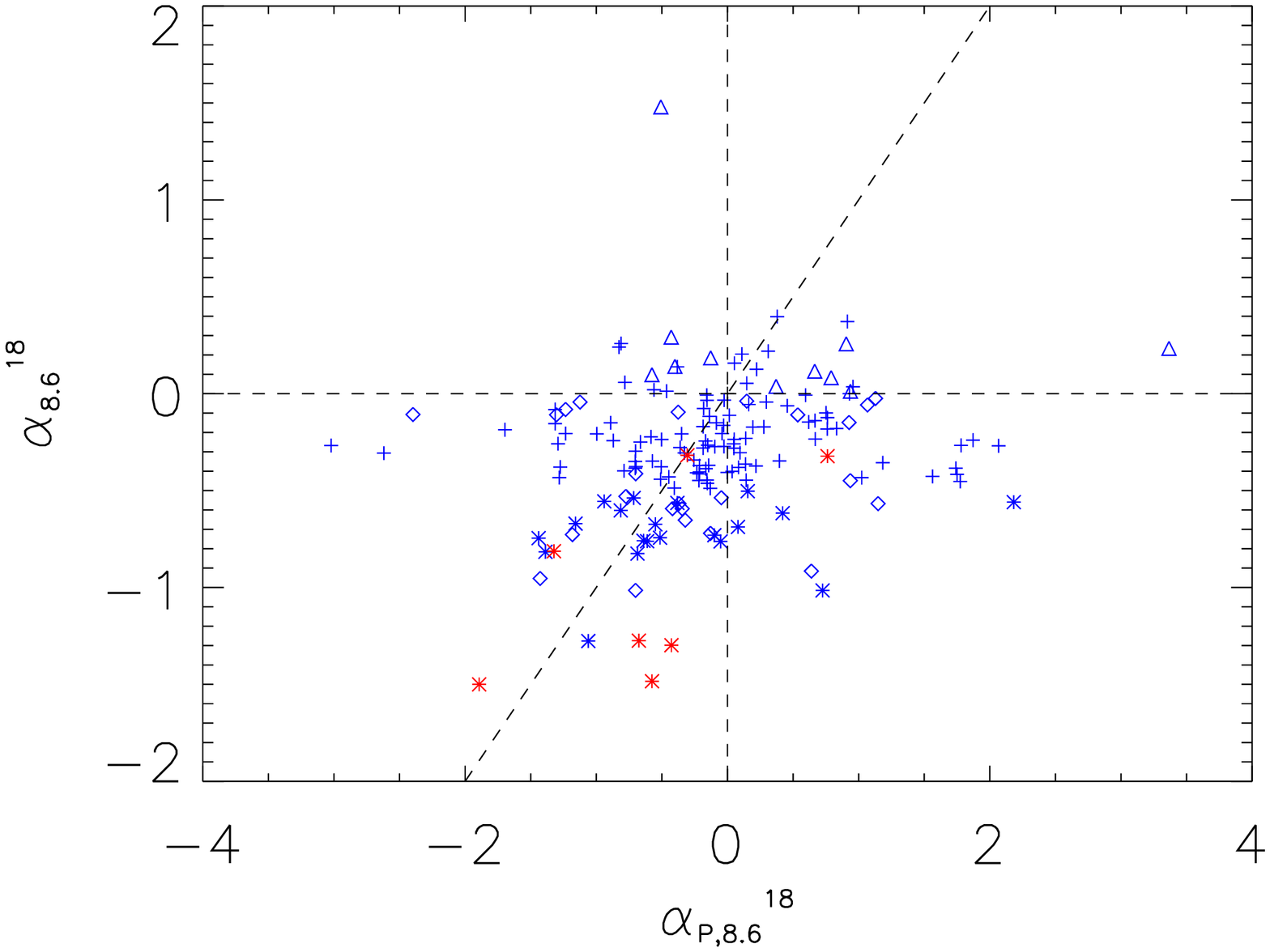}
 \caption{Comparison between the spectral index in total intensity and in polarisation in the ranges 4.8-8.6 GHz (upper panel) and 8.6-18 GHz.}
 \label{fig:alpha_PI}
\end{figure}

We have defined the spectral index $\alpha$ as $S \propto \nu^\alpha$. The analysis of AT20G data by Chhetri et al. (2012) has confirmed that $\alpha = -0.5$ is a physically meaningful threshold to separate compact, self-absorbed sources from structurally complex, extended objects. The majority (105, i.e. $61\%$) of our sources are flat spectrum objects (`F' $-0.5<\alpha_{4.8}^{8.6}, \alpha_{8.6}^{18}<0.5$). The remaining can be classified as follows: 25 ($15\%$) sources have peaked spectra (`P', $\alpha_{4.8}^{8.6}>0$ and $\alpha_{8.6}^{18}<0$); 29 ($17\%$) have steep spectra (`S', $\alpha_{4.8}^{8.6}<0$ and $\alpha_{8.6}^{18}<0$); 12 ($7\%$) have inverted (`I',$\alpha_{4.8}^{8.6}>0$ and $\alpha_{8.6}^{18}>0$)); only 1 source has upturning spectra (`U',$\alpha_{8.6}^{18}>0$).
Figure \ref{fig:cc} illustrates the spectral classification of the 172 sources in the sample with total intensity data at all the three frequencies (4.8, 8.6, and 18 GHz). The different distribution of the sources in the ($\alpha_{4.8}^{8.6}$, $\alpha_{8.6}^{18}<0.5$) plane indicates that the spectral properties over the 5 to 18 GHz frequency range may be different in total intensity and in polarisation (see also Fig. \ref{fig:alpha_PI}). The median values of the 4.8--8.6\,GHz and 8.6--18\,GHz spectral indices in polarisation are 0.20 and -0.16 respectively. This effect could be a combination of Faraday depolarisation operating at the lower frequencies, superposition of multiple components with different polarised spectra, and different magnetic field properties for the components that dominate the emission at the different frequencies. 

We constructed the matrix in Table \ref{Table:matrix} by comparing the total intensity and the polarisation spectral behaviours. The distribution across the cells confirms the differences of spectral behaviour in polarisation and in total intensity. Similar behaviours translate in a diagonal matrix. However, even if a tiny effect due to Faraday depolarisation could be the cause of the high number of peaked spectra in polarised emission (because of the lower level of emission at lower frequencies), more than $\sim 30 \%$ of sources show a polarised emission spectral index on the lower frequency range higher than at the higher frequencies, indicating that Faraday emission is not significantly affecting their spectral behaviour.

Chhetri et al. (2012) demonstrate that low frequency spectral index selections in flat and steep populations are more effective in identifying compact and extended objects than high frequency spectral indices. For this reason we rely on the $\alpha_{4.8}^{8.6}$ to classify flat and steep spectra sources in some of the following analysis. We found 163 flat spectra objects and only 9 steep spectra objects.
Figure \ref{fig:cc} also shows that for this selection criterion the total intensity and polarisation spectral behaviours are different.
The median of spectral indices in total intensity and polarised flux densities for each class and for the full sample are in Table \ref{Table:spind}. An overall steepening is appreciable both in total intensity and in polarised emission.

\subsection{Fractional polarisation}

\begin{figure}

\includegraphics[trim=3cm 0cm 0cm 0cm, width=9.5cm, clip=true]{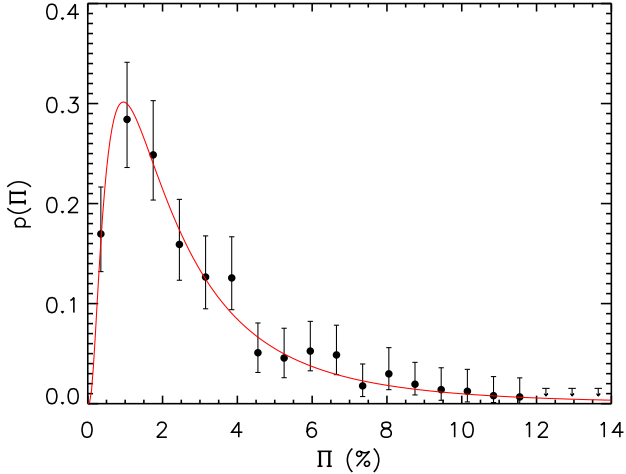}
\caption{Distributions of the polarisation degree at 18 GHz. Errors and upper limits corresponds to a 68\% c.l.. The solid line shows the log-normal distribution with median fractional polarisation 2.14$\%$ and $\sigma=0.90\%$ (see the text for details).}
 \label{fig:pol_deg_histo}
\end{figure}

We derived the distribution of the polarisation fractions in Fig. \ref{fig:pol_deg_histo} and Table \ref{Table:poldeg} using a bootstrap re-sampling method in order to account for the uncertainties in both the polarised and the total intensity flux density measurements. 

We generated 1000 simulated catalogues by re-sampling, with repetitions, the input catalogue of polarised flux densities. In each simulation, values for the polarised and the total intensity flux densities were randomly assigned to each source by assuming a Gaussian distribution with a mean equal to the measured values and a $\sigma$ equal to the quoted errors. When only an upper limit was available on the polarised flux density, we generated random values between 0 and the quoted upper limit assuming a uniform distribution. For each realization, the polarisation fraction is estimated for each object as the ratio between the simulated polarised flux density and the simulated total intensity flux density. The resulting values are then distributed into
bins of polarisation fraction. 

The final distribution of the polarisation fractions is given by the mean value of the simulated polarisation fractions in each bin with uncertainties derived assuming a Poisson statistic, according to the prescriptions of Gehrels (1986). In Fig. \ref{fig:pol_deg_histo} error bars correspond to the 68\% confidence interval.  

The solid line is a fit assuming the lognormal distribution:
$$
f(\Pi)= cost \cdot \frac{1}{\sqrt{2 \pi}\sigma \Pi} \ e^{(-0.5(\ln(\Pi/\Pi_m))^2/\sigma^2)},
$$
with $cost=0.97$, and $\sigma = 0.90\%$; $\Pi_m = 2.14\%$ is the median value of the distribution.

\begin{table}
 \caption{Distributions of the fractional polarisation at 18 GHz for the full sample.}
\centering
 \resizebox{6.2cm}{!}{
\begin{tabular}{cccc}
\hline
 $\Pi[\%]$ & Probability & $-$ & $+$\\
\hline
    0.35  &      0.170   &   0.038  &   0.047 \\
    1.05  &      0.284   &   0.048  &   0.057\\
    1.75  &      0.249   &   0.045  &   0.054\\
    2.45  &      0.159   &   0.036  &   0.045\\
    3.15  &      0.127   &   0.032  &   0.041\\
    3.85  &      0.126   &   0.032  &   0.041\\
    4.55  &      0.051   &   0.020  &   0.030\\
    5.25  &      0.046   &   0.020  &   0.030\\
    5.95  &      0.053   &   0.020  &   0.030\\
    6.65  &      0.049   &   0.020  &   0.030\\
    7.35  &      0.018   &   0.011  &   0.022\\
    8.05  &      0.030   &   0.016  &   0.026\\
    8.75  &      0.019   &   0.011  &   0.021\\
    9.45  &      0.014   &   0.011  &   0.021\\
   10.15  &      0.012   &   0.011  &   0.021\\
   10.85  &      0.008   &   0.007  &   0.019\\
   11.55  &      0.007   &   0.007  &   0.019\\
   12.25  &      $<$0.015  &          &\\
   12.95  &      $<$0.015  &          &\\
   13.65  &      $<$0.015  &          &\\
   14.35  &      $<$0.015   &          &\\
\hline
 \end{tabular}}
\label{Table:poldeg}
\end{table}

Table \ref{Table:frac_sp} summarizes the mean and quartiles of the distributions of fractional polarization  at 4.8, 8.6, and 18 GHz as calculated with survival analysis techniques to hold for upper limits in polarised flux density for each of the spectral classes identified considering the spectral behaviour in the ranges 4.8-8.6 and 8.6-18 GHz (see Fig. \ref{fig:cc}) as discussed in the previous section. While no significant trend is visible with frequency for flat, upturning or peaked spectra objects, there is a tiny indication that the mean polarisation fraction for steep spectrum sources decreases as the frequency decreases, probably under the effect of Faraday depolarisation, but the significance is poor because we deal with fairly broad distributions . 

\begin{table}
\centering
 \caption{Parameters describing the distributions of fractional polarisation at 4.8, 8.6, and 18 GHz for each spectral class defined according to Fig. \ref{fig:cc}. The spectral types are defined in the text (Upturning class has been neglected because it includes only 1 source).For each frequency and spectral class we quote the number of detections, the mean fractional polarisation and its error, the first, second and third quartiles of the distribution.} \label{Table:frac_sp}
 \resizebox{9cm}{!}{
\begin{tabular}{ccccc}
\hline
      & F & I  & P & S \\
\hline
18 GHz &&& \\
\hline
$N_{TOT}$   &  105 & 12  & 25 & 29 \\
Detections  &  101 & 11 & 25& 25 \\
$\langle \Pi \rangle {\pm \sigma_{\langle \Pi \rangle }}$ & 2.57 { $\pm$ 0.82} &1.55 {$\pm$ 0.25} &3.24 {$\pm$ 0.53}& 3.49 {$\pm$ 0.59} \\
{\tiny 1} 2 {\tiny 3} quartiles & {\tiny 1.21} 2.06 {\tiny 3.68} & {\tiny 1.00} 1.37 {\tiny 2.08} & {\tiny 1.13} 2.11 {\tiny 4.78} & {\tiny 0.25} 2.72 {\tiny 6.02}\\
\hline
8.6 GHz && \\
\hline
$N_{TOT}$   &  105 & 12  & 25 & 29 \\
Detections  &  100 & 10 & 24& 24 \\
$\langle \Pi \rangle {\pm \sigma_{\langle \Pi \rangle }}$ & 2.35 { $\pm$ 0.16} &1.28 {$\pm$ 0.29} &2.64 {$\pm$ 0.42}& 2.85 {$\pm$ 0.51} \\
{\tiny 1} 2 {\tiny 3} quartiles & {\tiny 1.10} 2.06 {\tiny 3.18} & {\tiny 0.30} 1.23 {\tiny 1.70} & {\tiny 1.07} 1.85 {\tiny 3.67} & {\tiny 0.17} 1.62 {\tiny 4.89}\\
\hline
4.8 GHz && \\
\hline
$N_{TOT}$   &  105 & 12  & 25 & 29 \\
Detections  &   93 & 10 & 24& 21 \\
$\langle \Pi \rangle {\pm \sigma_{\langle \Pi \rangle }}$ & 2.13 { $\pm$ 0.15} &1.48 {$\pm$ 0.30} &2.59 {$\pm$ 0.31}& 2.19 {$\pm$ 0.46} \\
{\tiny 1} 2 {\tiny 3} quartiles & {\tiny 0.95} 1.85 {\tiny 2.98} & {\tiny 0.55} 1.32 {\tiny 1.95} & {\tiny 1.40} 2.47 {\tiny 3.12} & {\tiny 0.00} 1.29 {\tiny 3.69}\\
\hline
\end{tabular}
}
\end{table}

\begin{table}
\centering
 \caption{Parameters describing the distributions of fractional polarisation for the full sample and for the flat- ($\alpha_{\nu1}^{\nu2}>-0.5$) and steep- ($\alpha_{\nu1}^{\nu2}<-0.5$) spectrum objects selected in the frequency ranges $\nu1=4.8, \nu2=8.6$ GHz. For each frequency and spectral class we quote the number of detections, the mean fractional polarisation and its error, the first, second and third quartiles of the distribution, and the probability that flat- and steep-spectrum objects are drawn from the same parent distribution, according to the two-sample Wilcoxon test.}
\label{Table:flatsteep}
 \resizebox{8.5cm}{!}{
 \begin{tabular}{cc|cc}
\hline
      & Full sample & Flat  & Steep \\
      &       & \multicolumn{2}{c}{selected between 4.8 and 8.6 GHz}\\
\hline
18 GHz &&& \\
\hline
$N_{TOT}$   &  187    & 163  & 9\\
Detections & 171 & 157 & 5\\
$\langle \Pi \rangle {\pm \sigma_{\langle \Pi \rangle }}$ & 2.79 { $\pm$ 0.17} &2.76 {$\pm$ 0.17} &2.32 {$\pm$ 0.99} \\
{\tiny 1} 2 {\tiny 3} quartiles & {\tiny 1.09} 2.04 {\tiny 3.84} & {\tiny 1.20} 2.08 {\tiny 3.82} & {\tiny 0.01} 0.55 {\tiny 2.54} \\
Prob(flat-steep)& &\multicolumn{2}{ c }{7.8\%}\\
\hline
8.6 GHz && \\
\hline
$N_{TOT}$   &  172     & 163  & 9 \\
Detections & 158 & 151 & 7\\
$\langle \Pi \rangle {\pm \sigma_{\langle \Pi \rangle }}$ &2.38 { $\pm$ 0.15} &2.42 {$\pm$ 0.15} & 1.80 { $\pm$ 0.77}\\
{\tiny 1} 2 {\tiny 3} quartiles & {\tiny 0.94} 1.80 {\tiny 3.27}  & {\tiny 1.03} 1.91 {\tiny 3.28} & {\tiny 0.04} 0.55 {\tiny 1.46}\\
Prob(flat-steep)&& \multicolumn{2}{ c }{5.3\%}\\
\hline
4.8 GHz && \\
\hline
$N_{TOT}$   & 173   & 164  & 9\\
Detections & 149 & 143 & 6\\
$\langle \Pi \rangle {\pm \sigma_{\langle \Pi \rangle }}$ &2.16 { $\pm$ 0.13} &2.21 { $\pm$ 0.13} &1.08 { $\pm$ 0.54}\\
{\tiny 1} 2 {\tiny 3} quartiles & {\tiny 0.87} 1.82 {\tiny 3.05} & {\tiny 0.94} 1.91 {\tiny 3.07} & {\tiny 0.00} 0.26 {\tiny 1.22} \\
Prob(flat-steep)& & \multicolumn{2}{ c }{0.1\%} \\
\hline
 \end{tabular}
 }
\end{table}

Table \ref{Table:flatsteep} shows the parameters describing the distributions of fractional polarisation for the full sample and for the flat- and steep-spectrum objects as calculated with survival analysis techniques to hold for upper limits in polarised flux density, selected according to the spectral behaviour in the 4.8-8.6 GHz frequency range only. There is a general trend towards an increase of the mean fractional polarisation as the frequency increases for the whole sample and for each spectral class (even if we have poor statistics for the steep-spectrum sources). This trend is not so noticeable if we consider the median values, similarly to the findings in Table \ref{Table:frac_sp}, as we deal with fairly broad distributions (see Table \ref{Table:flatsteep}). 

\begin{figure}
 \centering
 \includegraphics[trim=0cm 0cm 1cm 1cm, height=8cm, clip=true, angle=90]{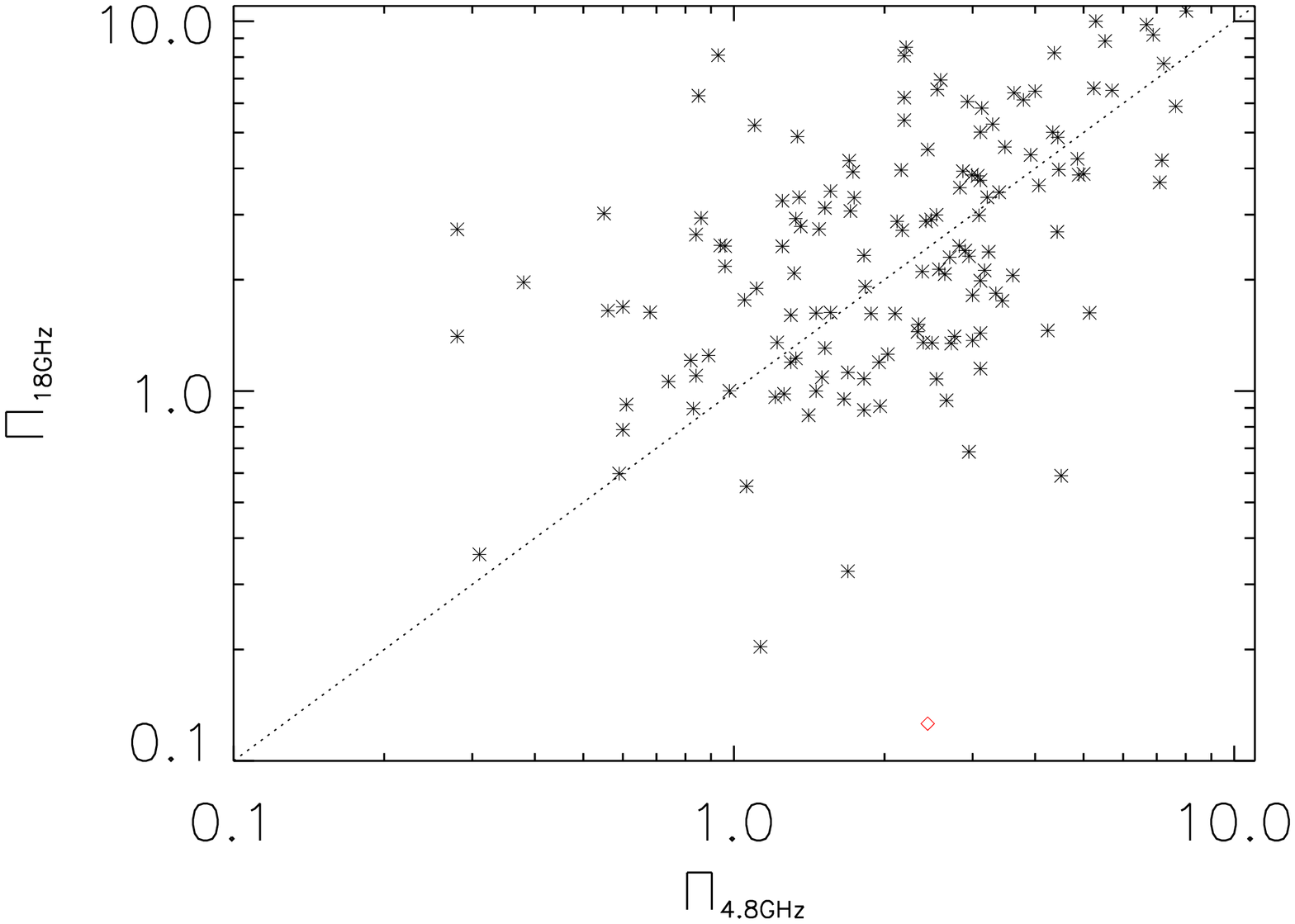}
 \includegraphics[trim=0cm 0cm 1cm 1cm, height=8cm, clip=true, angle=90]{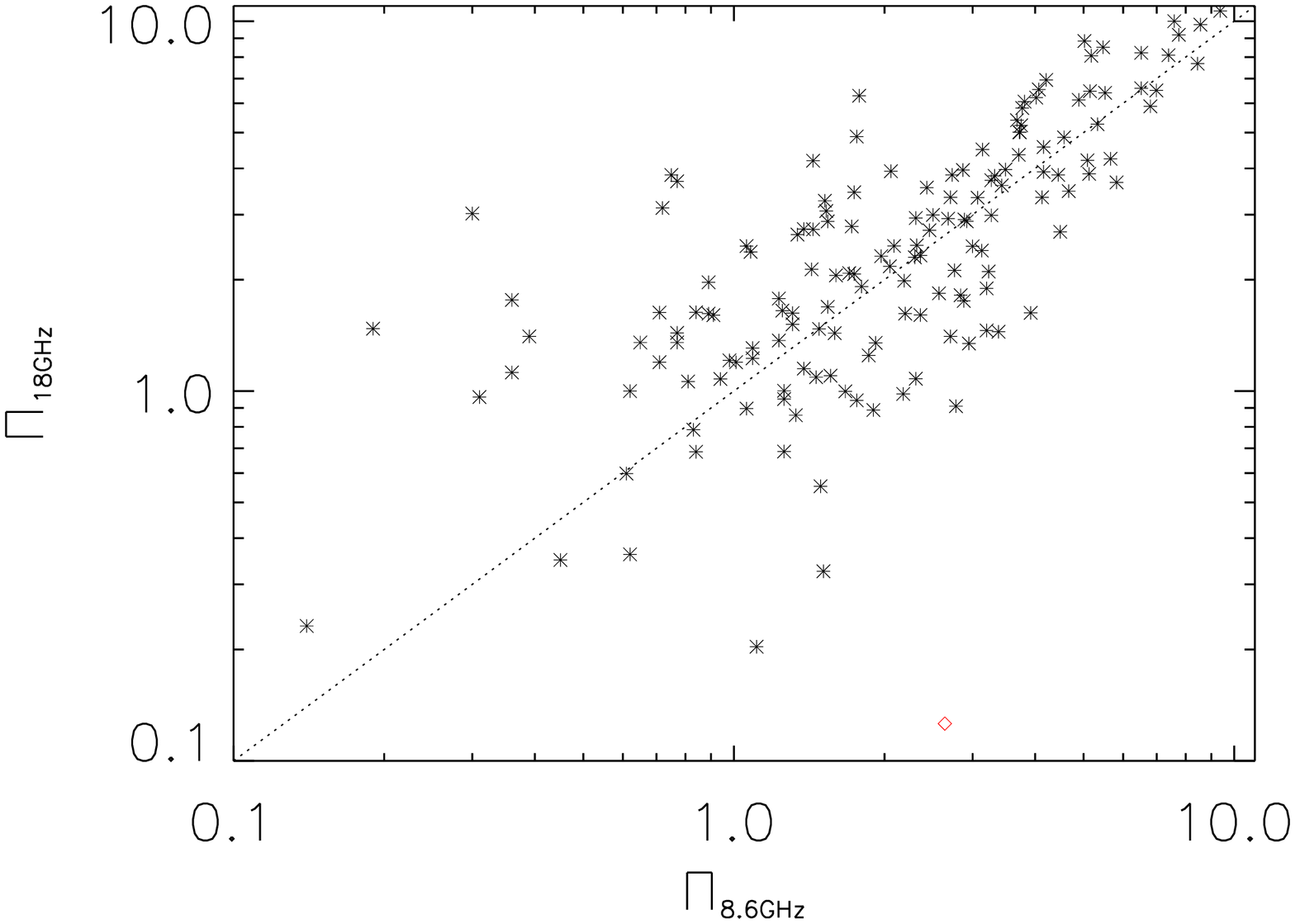}
 \caption{Fractional polarisation at 18 GHz as a function of 4.8 and 8.6 GHz for the full sample (red diamonds indicates upper limits at one or both the frequencies).}
 \label{fig:mm_mm}
\end{figure}

In fact, the differences among the polarisation degree distributions at different frequencies are not statistically significant. All the generalized Wilcoxon two-sample tests available in the ASURV package reject the null hypothesis, which states that the distributions of fractional polarisation are drawn from the same parent distribution, for the 4.8 and 8.6\,GHz distributions for the full sample. In the range 8.6-18 GHz a correlation is found by the different methods at $\sim 2-3 \sigma$ significance level. Between 4.8 and 18 GHz the Gehan's and Peto \& Peto version of the Wilcoxon test and the logrank test confirm a correlation at the $\sim 2 \sigma$ significance level, while the Peto \& Prentice version indicate a $\sim 4 \sigma$ level (see Fig. \ref{fig:mm_mm}). 

We then conclude that, for our sources, there is no statistically significant evidence of an increase of the median polarisation degree with a frequency above 4.8 GHz. This implies on the one hand that already at 4.8 GHz the Faraday depolarisation is not very important, and on the other hand that the magnetic field is not substantially more ordered in the regions dominating the emission at higher frequencies.

This conclusion is in line with the results by Battye et al. (2011) who found that the fractional polarisation of their sources, selected from the WMAP point source catalogue and mostly flat-spectrum, is almost independent of frequency in the range 8.4--43 GHz, with median values in the range 2--2.5\%. Klein et al. (2003) also found that flat-spectrum sources in their sample selected at 408 MHz are characterized by almost constant polarisation degrees, with median values $\simeq 2.5\%$, over the frequency range 2.7--10.5 GHz. On the other hand, they reported a steady increase of the polarisation degree of steep-spectrum sources from 2.2\% at 1.4 GHz to 5.8\% at 10.5 GHz. This increase, however, can be largely due to the bias induced by the requirement of polarisation detection at high frequencies: since steep-spectrum sources become faint at high frequencies, only those with exceptionally high polarisation degrees are detected. The same bias may explain the increase with frequency of the median polarisation degree observed by Sajina et al. (2011) for the subset of their AT20G sources with detected polarised flux in all four bands (4.86, 8.46, 22.46, and 43.34 GHz). This interpretation is consistent with their finding that the trend is stronger for steeper spectrum sources as well as for the lower flux density sources.

Several studies of radio source polarisation, mostly for samples selected at 1.4 GHz, dominated by steep-spectrum sources, have reported indications that the polarisation degree increases with decreasing flux density (Mesa et al. 2002; Tucci et al. 2004; Taylor et al. 2007; Grant et al. 2010; Subrahmanyan et al. 2010). A similar trend was also found by Sadler et al. (2006) for AT20G sources with $S_{20\rm GHz} > 100\,$mJy. It was, however, not confirmed by the analysis of the full AT20G sample by Massardi et al. (2011a), although the relatively low detection rate in polarisation prevented a clear conclusion. Our data, with a high detection rate, do not show any significant trend of $\Pi$ with flux density at any of the frequencies (4.8, 8.6 and 18 GHz) for $S_{18\rm GHz} > 500\,$mJy, with only 45\% probability that the null-hypothesis that sources with flux densities above and below 1 Jy at 18 GHz come from the same parent distribution is true. It must, however, be noted that our flux density range is limited and substantially narrower than that of Sadler et al. (2006).

The median of the ratio between the fractional polarisation at two different frequencies provides an estimate of the depolarisation as the frequency decreases. We found that $\langle \Pi_{4.8\rm GHz}/\Pi_{18\rm GHz}\rangle=79.7\%$ and $\langle \Pi_{8.6\rm GHz}/\Pi_{18\rm GHz}\rangle=86.9\%$, which implies a median depolarisation of 20.3\% and 13.1\% respectively in the two ranges of frequency 4.8-18 and 8.6-18 GHz. These findings confirm the results of Massardi et al. (2011a) and those by Tucci \& Toffolatti (2012).

\begin{table}
 \caption{Source counts at 18 GHz (including also WMAP data) as estimated for the current sample (upper panel) and for the whole AT20G sample (lower panel). See the text for details. Errors encloses the range of 68\% c.l..}
\centering
 \begin{tabular}{cccc}
  \hline
 $\log[P(Jy)]$   &  dN$/$dlogP $({\rm deg}^{-2}$)  & $-$    &  $+$\\
  \hline
         -1.428      &        0.0119       &    0.0026       &    0.0028\\
         -1.238     &         0.0086      &     0.0020      &     0.0027\\
         -1.047     &         0.0042      &     0.0015      &     0.0019\\
         -0.857     &         0.0013      &     0.0006      &     0.0017\\
         -0.666     &         0.0014      &     0.0007      &     0.0016\\
         -0.476      &        0.0005      &     0.0004      &     0.0012\\
         -0.285      &        0.0005      &     0.0004      &     0.0012\\
         -0.095      &        $<$0.0009&&\\
          0.096      &        0.0005  &         0.0004      &     0.0012\\
          0.287      &        $<$0.0009&&\\
          0.477      &        0.0005  &         0.0004      &     0.0012\\
 
  \hline   
         -2.000         &     0.1115      &     0.0055     &      0.0053\\
         -1.809         &     0.0624      &     0.0040     &      0.0041\\
         -1.619         &     0.0353      &     0.0030     &      0.0031\\
         -1.428         &     0.0203      &     0.0024     &      0.0025\\
         -1.238         &     0.0114      &     0.0017     &      0.0021\\
         -1.047         &     0.0062      &     0.0012     &      0.0017\\
         -0.857         &     0.0032      &     0.0010     &      0.0011\\
         -0.666         &     0.0016      &     0.0006     &      0.0009\\
         -0.476         &     0.0008      &     0.0004     &      0.0008\\
         -0.285         &     0.0003      &     0.0003     &      0.0005\\
         -0.095         &     $<$0.0004&&\\
          0.096         &     $<$0.0004&&\\
          0.287         &     $<$0.0004&&\\
          0.477         &     $<$0.0004&&\\
\hline 
\end{tabular}
\label{Table:sourcecounts}
\end{table}
\section{Source counts in polarisation}\label{sec:counts}

\begin{figure}
 \includegraphics[trim=3cm 0cm 0cm 0cm, width=8cm, clip=true]{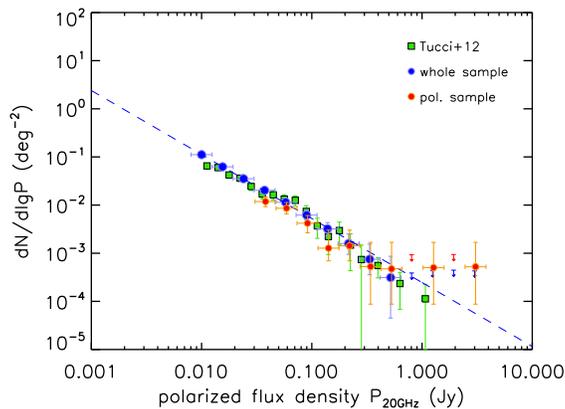}
 \caption{18 GHz differential source counts in polarisation calculated with the two methods described in the text compared with the findings by Tucci et al. (2012). The dashed line represent a linear fit of the whole AT20G sample data.}
 \label{fig:srccnt}
\end{figure}

We derived the differential number counts in the polarised flux density in two ways. Results are listed in Table \ref{Table:sourcecounts} and shown in Fig. \ref{fig:srccnt}. In both cases we used the bootstrap resampling method and performed 1000 simulations. 

As a first method, we directly measured the number counts from the catalogue of polarised flux density measurements. For each simulation, the catalogue was resampled with repetitions. As before, a value for each of the polarised flux density was randomly assigned to each source by assuming a Gaussian distribution with a mean equal to the measured values and $\sigma$ equal to the quoted errors. When only an upper limit was available on the polarised flux density, we generated random values between 0 and the quoted upper limit assuming a uniform distribution. The distribution of the polarised flux densities derived from the simulations were binned into a histogram, and the mean value in each bin was taken as the measurement of the number count in that bin, after dividing by the bin size and by the survey area. The errors on the number counts were derived assuming a Poisson statistic, according to the prescriptions of Gehrels (1986). The results are shown in Fig. \ref{fig:srccnt} by the red dots with error bars corresponding to the 68\% confidence interval.

For the second method we started off with the Massardi et al. (2011a) catalogue of total intensity flux densities. In each simulation the total intensity flux densities was randomly assigned to each source by assuming a Gaussian distribution with a mean equal to the measured values and sigma equal to the quoted errors. At the same time a realization of the polarisation fraction was produced from the catalogue of polarised flux densities following the procedure described before. The simulated total intensity flux density and the simulated value of the polarisation fraction were then used to derive the corresponding polarised flux density of each object. Finally the distribution of the polarised flux densities obtained from the simulations were binned into a histogram and the mean value in each bin was taken as the measurement of the number count in that bin, divided by the bin size and by the survey area. The errors on the number counts were derived assuming a Poisson statistic, according to the prescriptions of Gehrels (1986). The results are shown in Fig. \ref{fig:srccnt} by the blue dots with error bars corresponding to the 68\% confidence interval. These points are linearly fitted with the relation:
$$
\log (dN /d\log P [deg^{-2}]) = (-1.32 \pm 0.05) \log P[Jy] + (-3.60 \pm 0.09).
$$

In the same Figure the green squares represent the number counts in the polarised flux densities derived by Tucci \& Toffolatti (2012).

\section{Polarisation calibrators at millimetric wavelengths}\label{sec:cal}

\begin{table*}
\caption{Candidate millimetric polarisation calibrators (see the text for details on selection criteria). Columns are as follows: (1) AT20G name; (2) 18 GHz total intensity flux density and its error in mJy; (3) 18 GHz fractional polarisation; (4-5) spectral indices $\alpha_{4.8}^{8.6}$ and $\alpha_{8.6}^{18}$ where low frequency AT20G data are available; (6) variations of fractional polarisation at different frequencies $\Delta_\Pi = \Pi_{8.6 GHz}-\Pi_{18 GHz}$ where low frequency AT20G data are available; (7-8) epoch of AT20G catalogue run (see Murphy et al. 2010; `1': 2004, `2': 2005,`3-4': 2006,`5-6': 2007,`7': 2008) and relative variation of flux densities $(S_{AT20G}-S_{Oct2006})/S_{AT20G}$; (9) ratio between the visibilities amplitude averaged over the long baselines ($\sim4.5$ km) and the short ones ($\sim0.2$ km) in the AT20G catalogue (Chhetri et al. 2013); (10) 3~mm flux density extracted from the ATCA calibrator database; (11-14) flux densities listed (above $5\sigma$) in the {\sc Planck} Legacy Catalogue (Planck Collaboration, 2013) at 30, 100, 217, and 353 GHz, in mJy; (15-16) redshift and optical identification (`QSO-G': quasar and galaxy as classified in NED,`AeB': AGN with broad line emissions, `Ae': AGN with emission features, Mahony et al. 2011).}\label{Table:calib}			
\resizebox{18cm}{!}{
\begin{tabular}{lllccccccccccccc}
\hline																			
Name `AT20GJ'	&	S20		&	m20	&	$\alpha_{4.8}^{8.6}$	&	$\alpha_{8.6}^{20}$		&	$\Delta_\Pi$	&	Run &Variab	&	6kmVis	&	S\_ATCAcal	&	S\_ {\sc Planck}	&	S\_ {\sc Planck} &	S\_ {\sc Planck}	&	S\_ {\sc Planck}	&	z	&	Optid	\\
	 	&	 		[mJy]		&	[\%]	&		&		&		[\%]	&		& 18 GHz		& ratio		&	3mm[mJy]	&	30GHz	&	100GHz	&	217GHz&	353GHz	&		&		\\
\hline																			
    J001035-302748	&		595.8$\pm$11.9	&	5.23	&	0.26	&	-0.17	&		-1.48	&	1	&	-24.37	&	0.96	&	...	&	...	&	...	&	...	&	...	&	1.19	&	QSO	\\
	J004959-573827	&		1859.0$\pm$37.2	&	3.84	&	0.16	&	-0.38	&		-1.11	&	2	&	-0.70	&	0.97	&	621.3	&	1262.19	&	651.72	&	324.88	&	...	&	1.797	&	QSO	\\
	J040658-382627	&		1627.0$\pm$32.6	&	6.13	&	0.07	&	-0.26	&		-1.24	&	1	&	28.15	&	0.94	&	786.5	&	1402.08	&	925.25	&	552.50	&	374.80	&	1.285	&	QSO	\\
	J045550-461558	&		5333.0$\pm$106.7	&	6.54	&	0.60	&	-0.11	&		-2.47	&	1	&	22.05	&	...	&	1204.6	&	2596.76	&	1407.34	&	840.84	&	516.70	&	0.853	&	QSO	\\
	J051644-620706	&		645.7$\pm$12.9	&	4.57	&	0.14	&	-0.11	&		-0.41	&	2	&	-22.81	&	0.98	&	...	&	1183.03	&	633.39	&	630.04	&	379.25	&	...	&	BLLac	\\
	J055009-573224	&		976.7$\pm$19.5	&	5.27	&	0.36	&	-0.17	&		0.06	&	2	&	-2.90	&	0.99	&	...	&	1171.74	&	818.22	&	599.80	&	505.13	&	2.001	&	QSO	\\
	J062307-643620	&		513.1$\pm$10.3	&	3.33	&	-0.14	&	-0.26	&		-0.26	&	2	&	-39.74	&	0.97	&	432.4	&	835.80	&	546.79	&	393.24	&	390.74	&	0.129	&	G	\\
	J063546-751616	&		3311.0$\pm$66.3	&	6.21	&	-0.17	&	-0.37	&		-2.19	&	4	&	5.10	&	0.99	&	2618	&	4295.48	&	3406.59	&	1831.33	&	1179.58	&	0.653	&	QSO	\\
	J074719-331046	&	1105.0$\pm$22.1	&	3.55	&	0.39	&	0.37	&		-1.12	&	1	&	13.94	&	...	&	1283.7	&	...	&	1035.80	&	884.28	&		&	...	&	...	\\
	J081108-492943	&		544.9$\pm$10.9	&	3.47	&	0.03	&	-0.30	&		1.2	&	1	&	-12.31	&	...	&	...	&	...	&	...	&	...	&	...	&	...	&	...	\\
	J102343-664647	&		519.9$\pm$10.4	&	3.84	&	-0.16	&	-0.45	&		-3.09	&	2	&	-6.37	&	0.96	&	...	&	...	&	248.99	&		&	...	&	...	&	...	\\
	J110352-535700	&		555.8$\pm$11.1	&	6.29	&	0.44	&	-0.57	&		-4.51	&	2	&	3.02	&	0.94	&	...	&	684.57	&	871.58	&	545.33	&	...	&	...	&	...	\\
	J111207-570339	&		643.4$\pm$12.9	&	8.1	&	-0.02	&	-0.67	&		-0.71	&	2	&	-15.17	&	...	&	...	&		&	400.88	&	300.47	&	...	&	...	&	...	\\
	J111827-463415	&		857.6$\pm$17.2	&	3.67	&	-0.02	&	-0.24	&		2.15	&	1	&	-11.59	&	...	&	...	&	1034.80	&	769.83	&	404.30	&	...	&	0.713	&	 QSO 	\\
	J115253-834410	&		521.0$\pm$24.0	&	6.50	&	0.46	&	0.16	&		0.49	&	3	&	0.00	&	1	&	...	&	...	&	321.68	&	...	&	...	&	...	&	...	\\
	J121806-460030	&		687.6$\pm$13.8	&	8.50	&	-0.55	&	-1.27	&		-3.03	&	1	&	-1.22	&	0.89	&	...	&	...	&	...	&	...	&	...	&	0.529	&	QSO	\\
	J142432-491349	&		2635.0$\pm$52.8	&	6.94	&	-0.36	&	-0.76	&		-2.73	&	5	&	-0.27	&	0.76	&	1226	&	1942.63	&	992.69	&	616.51	&		&	...	&	...	\\
	J142756-420618	&		2495.0$\pm$49.9	&	3.96	&	0.04	&	-0.40	&		-1.09	&	5	&	-9.98	&	0.98	&	980.8	&	1972.67	&	1898.62	&	1332.41	&	868.12	&	1.522	&	QSO	\\
	J170144-562155	&	1123.0$\pm$22.5	&	6.06	&	-0.16	&	-0.73	&		-2.25	&	5	&	-3.38	&	1.01	&	500.7	&	1074.82	&	615.90	&	283.17	&	...	&	...	&	...	\\
	J170918-352520	&		568.9$\pm$11.4	&	8.84	&	-0.31	&	-0.69	&		-3.82	&	1	&	-6.87	&	0.53	&	...	&	...	&	...	&	...	&	...	&	...	&	...	\\
	J171310-341827	&		908.2$\pm$18.2	&	4.85	&	0.32	&	-0.49	&		-0.28	&	1	&	-12.97	&	1.01	&	...	&	...	&	...	&	...	&	...	&	...	&	...	\\
	J173315-372232	&		1001.0$\pm$20.0	&	3.45	&	0.49	&	0.00	&		-1.71	&	5	&	-12.19	&	1.02	&	1270.6	&	...	&	1426.47	&	955.66	&	...	&	...	&	...	\\
	J182607-365102	&	558.6$\pm$11.2	&	3.60	&	0.34	&	-0.15	&		-0.17	&	4	&	4.40	&	1.01	&	...	&	...	&	...	&	...	&	...	&	...	&	...	\\
	J194025-690757	&		596.7$\pm$12.0	&	3.13	&	-0.13	&	-0.43	&		-2.41	&	2	&	12.52	&	0.98	&	...	&	...	&	...	&	...	&	...	&	3.154	&	QSO	\\
	J210933-411020	&		1905.0$\pm$38.1	&	10.01	&	0.06	&	-0.49	&		-2.42	&	1	&	14.38	&	...	&	790.2	&	698.88	&	428.23	&	182.42	&	...	&	1.058	&	QSO	\\
	J214629-775554	&		1264.0$\pm$2.0	&	3.02	&	0.91	&	0.23	&		-2.72	&	4	&	0.00	&	0.99	&	...	&	...	&	399.99	&	238.80	&	...	&	0.334	&	QSO	\\
	J220743-534633	&		1133.0$\pm$22.7	&	4.25	&	-0.06	&	-0.40	&		1.41	&	2	&	0.97	&	0.95	&	...	&	...	&	548.39	&	305.55	&	...	&	1.206	&	QSO	\\
	J224616-560746	&		749.7$\pm$15.0	&	8.07	&	0.43	&	-0.01	&		-2.89	&	2	&	-40.32	&	1	&	...	&	...	&	271.68	&	154.80	&	...	&	1.325	&	QSO	\\
	J232917-473019	&	1153.0$\pm$23.1	&	3.97	&	-0.23	&	-0.41	&		-0.48	&	1	&	-23.07	&	0.96	&	1023.3	&	1578.01	&	1050.37	&	626.79	&	436.81	&	1.304	&	AeB	\\
 \hline
  \end{tabular}
  }
 \end{table*}
 
Finding suitable calibrators for polarisation studies at millimetric wavelengths is a difficult exercise. On the one hand, extragalactic sources typically have a very low level of circular polarisation, which simplifies the calibration solutions. On the other hand, as demonstrated in the previous sections, the fractional polarisation is typically a small fraction of the total intensity and polarised behaviour cannot be easily predicted on the basis of total intensity properties. Furthermore, source variability makes the fractional polarisation unknown. Source compactness (with respect to the beam of the used telescope) allows us to consider only on-axis effects. Models of extended emissions of calibrators could be used to cope with off-axis effects, if there is no better suited compact calibrator.

Polarised sources allow to recover gain and instrumental polarisation parameters calibration, even if the polarisation fraction of the calibrator is unknown (but non-zero) and its polarisation angle is unknown, by observing over a wide range of parallactic angles (for alt-az telescope mounts and assuming the instrumental parameters do not vary in the time of the observation, see Sault, Hamaker \& Bregman 1996).  

Statistical analysis clearly showed that spectral behaviours in total intensity and polarisation are typically different, with a tiny indication that the fractional polarisation increases with frequency, at least for high-frequency-selected steep-spectra sources. Therefore, it is difficult to make predictions with our data at frequencies $\gsim 20$ GHz. Any candidate selection requires proper monitoring at the observing frequency to confirm the calibrator properties. Hence, we stress the fact that in this section we do not claim any definition of effective criteria to identify good calibrators, but we only suggest a short list of targets for future calibrator monitoring programs at frequencies $\gsim 20$ GHz . 

The list is selected in the catalogue presented in the previous sections of sources
\begin{itemize}
\item south of $-30^\circ$ deg, excluding the Galactic plane region ($|b|\le 1.5^\circ$) and the LMC region
\item with total intensity flux density at 20 GHz $ S_{20 GHz}>500$ mJy in the 2006 AT20G selection. 
\end{itemize}

Among them we selected the sources with fractional polarisation at 20 GHz $\Pi_{20 GHz}>3$\% (corresponding to $P>15$ mJy and, according to the distribution in Sect. \ref{sec:analysis}, enclosing about $40\%$ of our sources) and at least two of the following properties
\begin{itemize}
\item flat spectral index between 8 and 20 GHz $\alpha_8^{20}>-0.5$ in the 2006 epoch, to enhance the chances of choosing sources that are bright above 20 GHz;
\item $S_{20 GHz}>500$ mJy in all the AT20G epochs, and varying less than 10\% with respect to the October 2006 epoch over 1 yr of observations or smaller than 20\% over longer epochs to select the most stable objects according to the available data;
\item increasing fractional polarisation between 8 and 20 GHz, to favour high levels of fractional polarisation observations at the higher frequencies.
\end{itemize}

The selected sample contains 29 sources. They have been flagged in Table \ref{Table:fullTable} with a `c' and some of their properties are summarized in Table \ref{Table:calib}. All the sources are classified as point-like in the AT20G catalogue (i.e. with most of their emission within the synthesized beam of 10 arcsec at 20 GHz). 

Furthermore, where available, we considered the source `6~km visibility' which is the ratio of the scalar averaged amplitudes at the long baselines ($\sim$4.5~km) over the short baselines ($\sim$0.2 km) of the AT20G catalogue. Chhetri et al. (2013) provided an efficient tool to select point like sources in the AT20G, where 6~km visibility values larger than 0.86 at 20 GHz identify sources with angular size smaller than 0.15 arcsec, which makes them good calibrator candidates. Only two sources (AT20GJ142432-491349, AT20GJ170918-352520) have ratios smaller than the threshold, but they could still be candidates for observations with synthesized beams on the scale of $\gsim 1$ arcsec.

As expected, spectral behaviour are mostly slowly steepening or peaked. We also note that Chhetri et al. (2012) found that compact sources show spectral steepening at rest frequencies $\sim30$ GHz, which may affect the higher frequency spectra. Optical identifications (Mahony et al. 2011) indicate that the sample is mostly composed by QSO at mean redshift above 1.2. Almost all the objects are calibrators in the ATCA database and the flux densities at 3~mm available for 10 of them are larger than 400 mJy. Detections with total intensity $>5\sigma$ are available in the {\sc Planck} Legacy Catalogue (Planck Collaboration 2013) at 100,  217, and 353 GHz channels  (roughly corresponding to ALMA Band 3, 6, and 7 frequencies) in the position of 22 sources with median values above 450 mJy for all the bands. The median fractional polarisation of the sample is above 4.8\%. The most polarised object is AT20GJ210933-411020 with 1.9~Jy of total intensity and 10\% fractional polarisation, flat spectrum, increasing fractional polarisation with frequency, and only 14\% relative variability over 3~yr time. In the {\sc Planck} channels the spectra become steep down to 428 mJy at 100 GHz and only 128 mJy at 217 GHz.

AT20GJ063546-751616 has 5.33 Jy of total intensity flux density at 20 GHz, 6.2\% polarised, and remains above 1 Jy up to 1~mm frequencies. It is classified as a flat-spectrum radio quasar at redshift z=0.653. Several notes indicate the presence of a jet structure, but the 6 km visibilities identify it as point-like and modestly variable in ATCA observations over few years. Thanks to its position it is always visible to Southern hemisphere telescopes like ATCA and ALMA and stands as the most suitable polarisation calibrator at high frequencies and low declinations.

\section{Conclusions} \label{sec:conclusions}
We have conducted sensitive polarisation and total intensity observations on a 20\,GHz flux-limited sample of 189 objects selected in the Australia Telescope 20\,GHz Survey, choosing sources that have $S_{\rm 20 GHz}>500$ mJy in the declination range $\delta<-30^\circ$ in the survey scans before October 2006. They have been followed up during an observing run in October 2006 designed to reach 1\,mJy sensitivity in polarisation. This strongly improved the sensitivity and the detection rate for polarisation observations over any previous sample investigated in this sky area at frequencies above 10\,GHz.

94\% of the 180 extragalactic point sources have a detection of polarised flux density at least at 18 GHz. 172 of them have been observed also at 4.8 and 8.6 GHz, and 143 sources have a detected polarised flux density at all three frequencies.

The 9 sources identified as extended have poor quality flux density measurement. So, for the sake of completeness, we extracted the values of polarised flux density from the 9-yr co-added WMAP maps. We recover an upper limit for 5 of them and a detection at 23 GHz for 2 of them (ForA and CenA). The final sample of 187 sources that we analyzed constitutes a 99\% complete sample at the 2006 survey selection epoch with a 91.4\% polarisation detection rate. In addition, detections have been obtained at all the WMAP frequencies for PicA.

This sample constitutes an ancillary data set for present and future studies of polarisation in the Southern Hemisphere and complements other samples recently observed either in equatorial regions (Sajina et al. 2011) or in the Northern hemisphere (Jackson et al. 2010). Analysis of the WMAP and  {\sc Planck} data (L\'opez-Caniego et al. 2009 and references therein) has demonstrated that similar source lists are crucial to improve the investigation of the CMB E and B modes in millimetric wavelength bands.

Thanks to our high detection rate, to a low polarised flux density level, to the multi-frequency observations, and to the inclusion of integrated flux densities for extended objects observed in mosaic mode with the ATCA (Burke-Spolaor et al. 2009) or extracted from the WMAP 9-year maps (updating the findings of L\'opez-Caniego et al. 2009) the analysis of our sample in total intensity and polarisation allowed us to draw the following conclusions.
\begin{itemize}
	\item The spectral behaviours in total intensity and in polarisation are different for any population of sources. This implies that it is extremely difficult to make an estimation of polarised flux densities from total intensity measurements.
	\item There is no statistically significant evidence of increasing fractional polarisation with frequency. This implies that Faraday depolarisation is not strong enough to modify the spectral behaviour at and above $\sim4.8$ GHz. Spectral behaviour in polarisation is, in fact, the result of the combined effects of beam depolarisation due to multiple components, chaotic magnetic fields and Faraday depolarisation.
	\item Thanks to our high detection rate we can state that there is no evidence of an anticorrelation of fractional polarisation with total intensity flux density as was previously noted by several surveys, which were probably biased by a selection effect: only highly polarised sources can be detected for faint sources, while low fractional polarisation percentages can be detected in bright objects; furthermore faint objects in complete samples are typically more numerous than bright ones. 	
	\item Thanks to the high sensitivity of our observations we were able to extend the polarisation source counts at 18 GHz of Tucci \& Toffolatti (2012) and to confirm their findings. 
	\item We identified a list of 29 candidate calibrators for polarisation at declination below $-30^\circ$ and frequencies $\gsim 20$ GHz. The best candidate is AT20GJ063546-751616 that is point-like in the AT20G catalogue, has 5.33 Jy of total intensity flux density at 20 GHz, is 6.2\% polarised, and shows flux density above 1 Jy up to $\sim$1~mm wavelengths.
	
\end{itemize}

\section{Acknowledgements}
MM, MN, and GDZ acknowledges financial support for this research by ASI/INAF Agreement I/072/09/0 for the  {\sc Planck} LFI activity of Phase E2. ML-C acknowledges partial financial support from the Spanish Ministerio de Economía y Competitividad project AYA-2012-39475-C02-01 and the Consolider Ingenio-2010 Programme project CSD2010-00064.

We thank the staff at the Australia Telescope Compact Array site, Narrabri (NSW), for the valuable support they have provided in running the telescope. The Australia Telescope Compact Array is part of the Australia Telescope which is funded by the Commonwealth of Australia for operation as a National Facility managed by CSIRO. 

We thank Elisabetta Liuzzo (INAF-IRA) and Benjamin Walter (Haevrford) for the useful discussions and proofreading.
We thank the anonymous referee for the useful comments.


\begin{thebibliography}{}

\bibitem[\protect\citeauthoryear{Agudo et al.}{2010}]{2010ApJS..189....1A} Agudo I., Thum C., Wiesemeyer H., Krichbaum T.~P., 2010, ApJS, 189, 1


\bibitem[\protect\citeauthoryear{Arg{\"u}eso, Gonz{\'a}lez-Nuevo, \& Toffolatti}{2003}]{2003ApJ...598...86A} Arg{\"u}eso F., Gonz{\'a}lez-Nuevo J., Toffolatti L., 2003, ApJ, 598, 86


\bibitem[\protect\citeauthoryear{Argueso, Sanz, \& Herranz}{2011}]{2011arXiv1101.0701A} Argueso F., Sanz J.~L., Herranz D., 2011, arXiv, arXiv:1101.0701 

\bibitem[\protect\citeauthoryear{Battye et al.}{2011}]{2011MNRAS.413..132B} Battye R.~A., Browne I.~W.~A., Peel M.~W., Jackson N.~J., Dickinson C.,
2011, MNRAS, 413, 132

\bibitem[\protect\citeauthoryear{Blandford \& K\"onigl}{1979}]{1979ApJ...232...34B} Blandford R.~D., K{\"o}nigl A., 1979, ApJ, 232, 34

\bibitem[\protect\citeauthoryear{Burke-Spolaor et al.}{2009}]{2009MNRAS.395..504B} Burke-Spolaor S., Ekers R.~D., Massardi M., Murphy T., Partridge B., Ricci R., Sadler E.~M., 2009, MNRAS, 395, 504

\bibitem[\protect\citeauthoryear{Chhetri et al.}{2012}]{2012MNRAS.tmp.2747C} Chhetri R., Ekers R.~D., Mahony E.~K., Jones P.~A., Massardi M., Ricci R., Sadler E.~M., 2012, MNRAS, 2747

\bibitem[\protect\citeauthoryear{Chhetri et al.}{2013}]{2013arXiv1306.0990C} Chhetri R., Ekers R.~D., Jones P.~A., Ricci R., 2013, arXiv, arXiv:1306.0990 

\bibitem[\protect\citeauthoryear{Condon et al.}{1998}]{1998AJ....115.1693C} Condon J.~J., Cotton W.~D., Greisen E.~W., Yin Q.~F., Perley R.~A., Taylor G.~B., Broderick J.~J., 1998, AJ, 115, 1693

\bibitem[\protect\citeauthoryear{de Zotti et al.}{1999}]{1999AIPC..476..204D} de Zotti G., Toffolatti L., Arg{\"u}eso F., Davies R.~D., Mazzotta P., Partridge R.~B., Smoot G.~F., Vittorio N., 1999, AIPC, 476, 204

\bibitem[\protect\citeauthoryear{de Zotti et al.}{2005}]{2005A&A...431..893D} de Zotti G., Ricci R., Mesa D., Silva L., Mazzotta P., Toffolatti L., Gonz{\'a}lez-Nuevo J., 2005, A\&A, 431, 893


\bibitem[\protect\citeauthoryear{Gehrels}{1986}]{1986ApJ...303..336G} Gehrels N., 1986, ApJ, 303, 336 

\bibitem[\protect\citeauthoryear{Gold et al.}{2011}]{2011ApJS..192...15G} Gold B., et al., 2011, ApJS, 192, 15

\bibitem[\protect\citeauthoryear{Grant et al.}{2010}]{2010ApJ...714.1689G} Grant J.~K., Taylor A.~R., Stil J.~M., Landecker T.~L., Kothes R., Ransom R.~R., Scott D., 2010, ApJ, 714, 1689

\bibitem[\protect\citeauthoryear{Griffith et al.}{1994}]{gri94} Griffith M.R., Wright A.E., Burke B.F., Ekers R.D., 1994, ApJS, 90, 179

\bibitem[\protect\citeauthoryear{Griffith et al.}{1995}]{gri95} Griffith M.R., Wright A.E., Burke B.F., Ekers R.D., 1995, ApJS, 97, 347

\bibitem[\protect\citeauthoryear{Hancock et al.}{2011}]{2011ExA....32..147H} Hancock P.~J., et al., 2011, ExA, 32, 147


\bibitem[\protect\citeauthoryear{Jackson et al.}{2007}]{2007MNRAS.376..371J} Jackson N., Battye R.~A., Browne I.~W.~A., Joshi S., Muxlow T.~W.~B., Wilkinson P.~N., 2007, MNRAS, 376, 371

\bibitem[\protect\citeauthoryear{Jackson et al.}{2010}]{2010MNRAS.401.1388J} Jackson N., Browne I.~W.~A., Battye R.~A., Gabuzda D., Taylor A.~C., 2010, MNRAS, 401, 1388

\bibitem[\protect\citeauthoryear{Klein et al.}{2003}]{2003A&A...406..579K} Klein U., Mack K.-H., Gregorini L., Vigotti M., 2003, A\&A, 406, 579

\bibitem[\protect\citeauthoryear{Kurinsky et 
al.}{2013}]{2013A&A...549A.133K} Kurinsky N., Sajina A., Partridge B., Myers S., Chen X., L{\'o}pez-Caniego M., 2013, A\&A, 549, A133 



\bibitem[\protect\citeauthoryear{Mahony et al.}{2011}]{2011MNRAS.417.2651M} Mahony E.~K., et al., 2011, MNRAS, 417, 2651 

\bibitem[\protect\citeauthoryear{L{\'o}pez-Caniego et al.}{2009}]{2009ApJ...705..868L} L{\'o}pez-Caniego M., Massardi M., Gonz{\'a}lez-Nuevo J., Lanz L., Herranz D., De Zotti G., Sanz J.~L., Arg{\"u}eso F., 2009, ApJ, 705, 868

\bibitem[\protect\citeauthoryear{Massardi et al.}{2008}]{2008MNRAS.384..775M} Massardi M., et al., 2008, MNRAS, 384, 775

\bibitem[\protect\citeauthoryear{Massardi et al.}{2011a}]{2011MNRAS.412..318M} Massardi M., et al., 2011a, MNRAS, 412, 318

\bibitem[\protect\citeauthoryear{Massardi et al.}{2011b}]{2011MNRAS.415.1597M} Massardi M., Bonaldi A., Bonavera L., L{\'o}pez-Caniego M., de Zotti G., Ekers R.~D., 2011b, MNRAS, 415, 1597 

\bibitem[\protect\citeauthoryear{Mauch et al.}{2003}]{2003MNRAS.342.1117M} Mauch T., Murphy T., Buttery H.~J., Curran J., Hunstead R.~W., Piestrzynski B., Robertson J.~G., Sadler E.~M., 2003, MNRAS, 342, 1117


\bibitem[\protect\citeauthoryear{Mesa et al.}{2002}]{2002A&A...396..463M} Mesa D., Baccigalupi C., De Zotti G., Gregorini L., Mack K.-H., Vigotti M., Klein U., 2002, A\&A, 396, 463

\bibitem[\protect\citeauthoryear{Murphy et al.}{2010}]{2010MNRAS.402.2403M} Murphy T., et al., 2010, MNRAS, 402, 2403



\bibitem[\protect\citeauthoryear{ Planck}{2011a}]{2011A&A...536A...7P}  Planck, 2011a, A\&A, 536, A7

\bibitem[\protect\citeauthoryear{ Planck}{2011b}]{2011A&A...536A..13P}  Planck, 2011b, A\&A, 536, A13

\bibitem[\protect\citeauthoryear{ Planck et al.}{2013}]{2013arXiv1303.5088P}  Planck, et al., 2013, arXiv, arXiv:1303.5088 

\bibitem[\protect\citeauthoryear{Ricci et al.}{2004}]{2004A&A...415..549R} Ricci R., Prandoni I., Gruppioni C., Sault R.~J., De Zotti G., 2004, A\&A, 415, 549


\bibitem[\protect\citeauthoryear{Sadler et al.}{2006}]{2006MNRAS.371..898S} Sadler E.~M., et al., 2006, MNRAS, 371, 898

\bibitem[\protect\citeauthoryear{Sajina et al.}{2011}]{2011ApJ...732...45S} Sajina A., Partridge B., Evans T., Stefl S., Vechik N., Myers S., Dicker S., Korngut P., 2011, ApJ, 732, 45

\bibitem[\protect\citeauthoryear{Sault, Teuben, \& Wright}{1995}]{1995ASPC...77..433S} Sault R.~J., Teuben P.~J., Wright M.~C.~H., 1995, ASPC, 77, 433

\bibitem[\protect\citeauthoryear{Sault, Hamaker, 
\& Bregman}{1996}]{1996A&AS..117..149S} Sault R.~J., Hamaker J.~P., Bregman J.~D., 1996, A\&AS, 117, 149 

\bibitem[\protect\citeauthoryear{Sault, Rayner, \& Kesteven}{2002}]{2002AIPC..609..150S} Sault R.~J., Rayner D.~P., Kesteven M.~J., 2002, AIPC, 609, 150



\bibitem[\protect\citeauthoryear{Subrahmanyan et al.}{2010}]{2010MNRAS.402.2792S} Subrahmanyan R., Ekers R.~D., Saripalli L., Sadler E.~M., 2010, MNRAS, 402, 2792


\bibitem[\protect\citeauthoryear{Taylor et al.}{2007}]{2007ApJ...666..201T} Taylor A.~R., et al., 2007, ApJ, 666, 201

\bibitem[\protect\citeauthoryear{Toffolatti et al.}{1998}]{1998MNRAS.297..117T} Toffolatti L., Arg\"ueso Gomez F., de Zotti G., Mazzei P., Franceschini A., Danese L., Burigana C., 1998, MNRAS, 297, 117

\bibitem[\protect\citeauthoryear{Toffolatti et al.}{1999}]{1999ASPC..181..153T} Toffolatti L., De Zotti G., Arg{\"u}eso F., Burigana C., 1999, ASPC, 181, 153

\bibitem[\protect\citeauthoryear{Tucci et al.}{2004}]{Tucci04} Tucci M., Martinez-Gonzalez E., Toffolatti L., Gonzalez-Nuevo J., De Zotti G.; 2004; MNRAS, 349, 1267

\bibitem[\protect\citeauthoryear{Tucci \& Toffolatti}{2012}]{2012arXiv1204.0427T} Tucci M., Toffolatti L., 2012, arXiv:1204.0427



\end{thebibliography}
\end{document}